\documentclass[preprint2]{aastex6}
\usepackage{lineno}
\usepackage{xspace}

\usepackage{color}
\usepackage{appendix}
\definecolor{purple}{rgb}{0.62,0.12,0.94}
\newcommand\degrees{$^{\circ}$}
\newcommand{\aic}{\text{AIC}\xspace}

\begin{document}
\sloppy

\title{Search for extended sources in the Galactic Plane using 6 years of \emph{Fermi}-Large Area Telescope Pass 8 data above 10 GeV}

\author{
M.~Ackermann\altaffilmark{1}, 
M.~Ajello\altaffilmark{2}, 
L.~Baldini\altaffilmark{3}, 
J.~Ballet\altaffilmark{4}, 
G.~Barbiellini\altaffilmark{5,6}, 
D.~Bastieri\altaffilmark{7,8}, 
R.~Bellazzini\altaffilmark{9}, 
E.~Bissaldi\altaffilmark{10}, 
E.~D.~Bloom\altaffilmark{11}, 
R.~Bonino\altaffilmark{12,13}
% ...
}
\altaffiltext{1}{Deutsches Elektronen Synchrotron DESY, D-15738 Zeuthen, Germany}
\altaffiltext{2}{Department of Physics and Astronomy, Clemson University, Kinard Lab of Physics, Clemson, SC 29634-0978, USA}
\altaffiltext{3}{Universit\`a di Pisa and Istituto Nazionale di Fisica Nucleare, Sezione di Pisa I-56127 Pisa, Italy}
\altaffiltext{4}{Laboratoire AIM, CEA-IRFU/CNRS/Universit\'e Paris Diderot, Service d'Astrophysique, CEA Saclay, F-91191 Gif sur Yvette, France}
\altaffiltext{5}{Istituto Nazionale di Fisica Nucleare, Sezione di Trieste, I-34127 Trieste, Italy}
\altaffiltext{6}{Dipartimento di Fisica, Universit\`a di Trieste, I-34127 Trieste, Italy}
\altaffiltext{7}{Istituto Nazionale di Fisica Nucleare, Sezione di Padova, I-35131 Padova, Italy}
\altaffiltext{8}{Dipartimento di Fisica e Astronomia ``G. Galilei'', Universit\`a di Padova, I-35131 Padova, Italy}
\altaffiltext{9}{Istituto Nazionale di Fisica Nucleare, Sezione di Pisa, I-56127 Pisa, Italy}
\altaffiltext{10}{Istituto Nazionale di Fisica Nucleare, Sezione di Bari, I-70126 Bari, Italy}
\altaffiltext{11}{W. W. Hansen Experimental Physics Laboratory, Kavli Institute for Particle Astrophysics and Cosmology, Department of Physics and SLAC National Accelerator Laboratory, Stanford University, Stanford, CA 94305, USA}
\altaffiltext{12}{Istituto Nazionale di Fisica Nucleare, Sezione di Torino, I-10125 Torino, Italy}
\altaffiltext{13}{Dipartimento di Fisica, Universit\`a degli Studi di Torino, I-10125 Torino, Italy}
% ...

\begin{abstract}
The spatial extension of a $\gamma$-ray source is an essential ingredient to determine its spectral properties as well as its potential multi-wavelength counterpart. The capability to spatially resolve $\gamma$-ray sources is greatly improved by the newly delivered \emph{Fermi}-Large Area Telescope (LAT) Pass 8 event-level analysis which provides a greater acceptance and an improved point spread function, two crucial factors for the detection of extended sources. Here, we present a complete search for extended sources located within 7\degrees\ from the Galactic plane, using 6 years of \emph{Fermi}-LAT data above 10 GeV. We find 46 extended sources and provide their morphological and spectral characteristics. This constitutes the first catalog of hard \emph{Fermi}-LAT extended sources, named the Fermi Galactic Extended Source Catalog, which allows a thorough study of the properties of the Galactic plane in the sub-TeV domain.

\end{abstract}

\section{Introduction} \label{sec:intro}
Several surveys of the Galaxy have been undertaken at TeV $\gamma$-ray energies \citep[][for example]{2006ApJ...636..777A} by the current Instrument Atmospheric Cherenkov Telescopes (IACTs) revealing different classes of astrophysical sources such as supernova remnants (SNRs), pulsar wind nebulae (PWNe), and molecular clouds (MCs) \citep[][for a review on SNRs and PWNe]{2015CRPhy..16..674H}. Many are observed as spatially extended with respect to the angular resolution of the instruments. These sources produce $\gamma$-ray photons through inverse Compton (IC) scattering off highly relativistic leptons, bremsstrahlung radiation, or by hadrons interacting with interstellar matter. In many sources, this population of high energy particles emits GeV $\gamma$-rays detectable by the Large Area Telescope (LAT), the primary instrument on the \emph{Fermi} Gamma-Ray Space Telescope \citep{2009ApJ...697.1071A}. Indeed, since its launch in 2008, the \emph{Fermi}-LAT has detected a growing number of spatially extended sources across the sky thanks to its wide field of view ($\sim$2.4 sr) and (primarily) sky-survey operation mode. The Second \emph{Fermi}-LAT Point Source Catalog \citep[2FGL,][]{2012ApJS..199...31N} contained 12 extended sources. The number of extended sources increased to 22 in the First \emph{Fermi}-LAT Hard Source Catalog covering nearly 3 years of data in the range 10--500 GeV \citep[1FHL,][]{2013ApJS..209...34A}, to 25 in the Third \emph{Fermi}-LAT Point Source Catalog with 48 months of data in the range 100 MeV--300 GeV \citep[3FGL,][]{2015ApJS..218...23A} and to 31 in the Second \emph{Fermi}-LAT Hard Source Catalog with 80 months of data above 50 GeV \citep[2FHL,][]{2016ApJS..222....5A}. The addition of data and, in the case of the hard source catalogs, the focus on higher energies where photons are better localized and backgrounds are reduced have amplified the excellent capability of the \emph{Fermi}-LAT to spatially resolve GeV $\gamma$-ray sources.\\

Accurately estimating the spatial morphology of a $\gamma$-ray source is important for several reasons. Finding a coherent source extension across different energy bands can help to associate a \emph{Fermi}-LAT source with a potential counterpart. Such multi-wavelength studies can also help to determine the emission mechanisms producing these high energy photons. Due to the energy dependence of the \emph{Fermi}-LAT point spread function (PSF), the spatial and spectral characterization of a source cannot be decoupled. An incorrect spatial model will bias the spectral model of the source and vice versa, and can also skew the spectra of point sources in the vicinity of the extended source. \\

The 2FHL Catalog analyzed data from 50\,GeV to 2\,TeV and served to bridge the energy gap between ground-based $\gamma$-ray telescopes and the \emph{Fermi}-LAT. Of the 31 spatially extended sources found in 2FHL, 5 were detected as extended for the first time. The 2FHL showed that several of the extended sources previously identified by the \emph{Fermi}-LAT using lower energy data sets displayed a potential change in their best-fit extension and centroid, (i.e. the centroids and/or extensions of the 2FHL sources were not compatible within the errors to the corresponding 3FGL source).\\

In this paper we use 6 years of Pass~8 data to produce a catalog of extended sources detected by the \emph{Fermi}-LAT at energies between 10\,GeV and 2\,TeV at low Galactic latitude ($\pm$~$7^{\circ}$ of the Galactic plane). Lowering the energy threshold with respect to 2FHL to 10\,GeV maintains a PSF width $< 0.2^{\circ}$ and a reduced level of confusion from Galactic diffuse emission while increasing the number of $\gamma$-rays available for analysis. The lower energy threshold increases the number of detectable sources compared to 2FHL and permits a more robust measurement of morphology than 1) lower energy \emph{Fermi}-LAT data selections in regions where diffuse systematics are large and 2) higher energy \emph{Fermi}-LAT data selections for sources with fewer detected photons. This paper is the first catalog of extended sources produced with the \emph{Fermi}-LAT data, named the Fermi Galactic Extended Source (FGES) catalog, allowing a thorough study of the properties of the Galactic plane in the sub-TeV domain. The paper is organized as follows: Section~\ref{section:Description} describes the \emph{Fermi}-LAT and the observations used, Section~\ref{subsection:Analysis} presents our systematic methods for analyzing spatially extended \emph{Fermi}-LAT sources in the plane, Section~\ref{section:results} discusses the main results and a summary is provided in Section~\ref{sec:summary}.
\section{Fermi-LAT description and observations} 
\label{section:Description}

\subsection{\emph{Fermi}-LAT}
The \emph{Fermi}-LAT is a $\gamma$-ray telescope which detects photons by conversion into electron-positron pairs in the energy range between 20 MeV to higher than 500 GeV, as described in \cite{2009ApJ...697.1071A}. The LAT is composed of three primary detector subsystems: a high-resolution converter/tracker (for direction measurement of the incident $\gamma$-rays), a CsI(Tl) crystal calorimeter (for energy measurement), and an anti-coincidence detector to identify the background of charged particles. Since the launch of the spacecraft in June 2008, the LAT event-level analysis has been periodically upgraded
to take advantage of the increasing knowledge of how the \emph{Fermi}-LAT functions as well as the environment in which it operates. Following the Pass 7 data set, released in August 2011, Pass 8 is the latest version of the
\emph{Fermi}-LAT data. Its development is the result of a long-term effort aimed at a comprehensive revision of the entire event-level analysis and comes closer to realizing the full scientific potential of the \emph{Fermi}-LAT \citep{2013arXiv1303.3514A}. Compared to previous iterations of the \emph{Fermi}-LAT event-level analysis, Pass 8 provides a greater acceptance and an improved PSF\footnote{\url{http://www.slac.stanford.edu/exp/glast/groups/canda/lat_Performance.htm}} (with a 68~\% containment radius smaller than 0.2$^\circ$ above 10\,GeV that is nearly constant with increasing energy) which are two crucial factors for the detection of extended sources.

\subsection{Data selection}\label{subsection:DataSelection}
\begin{figure*}[!ht]
\begin{centering} 
\includegraphics[width=0.98\textwidth]{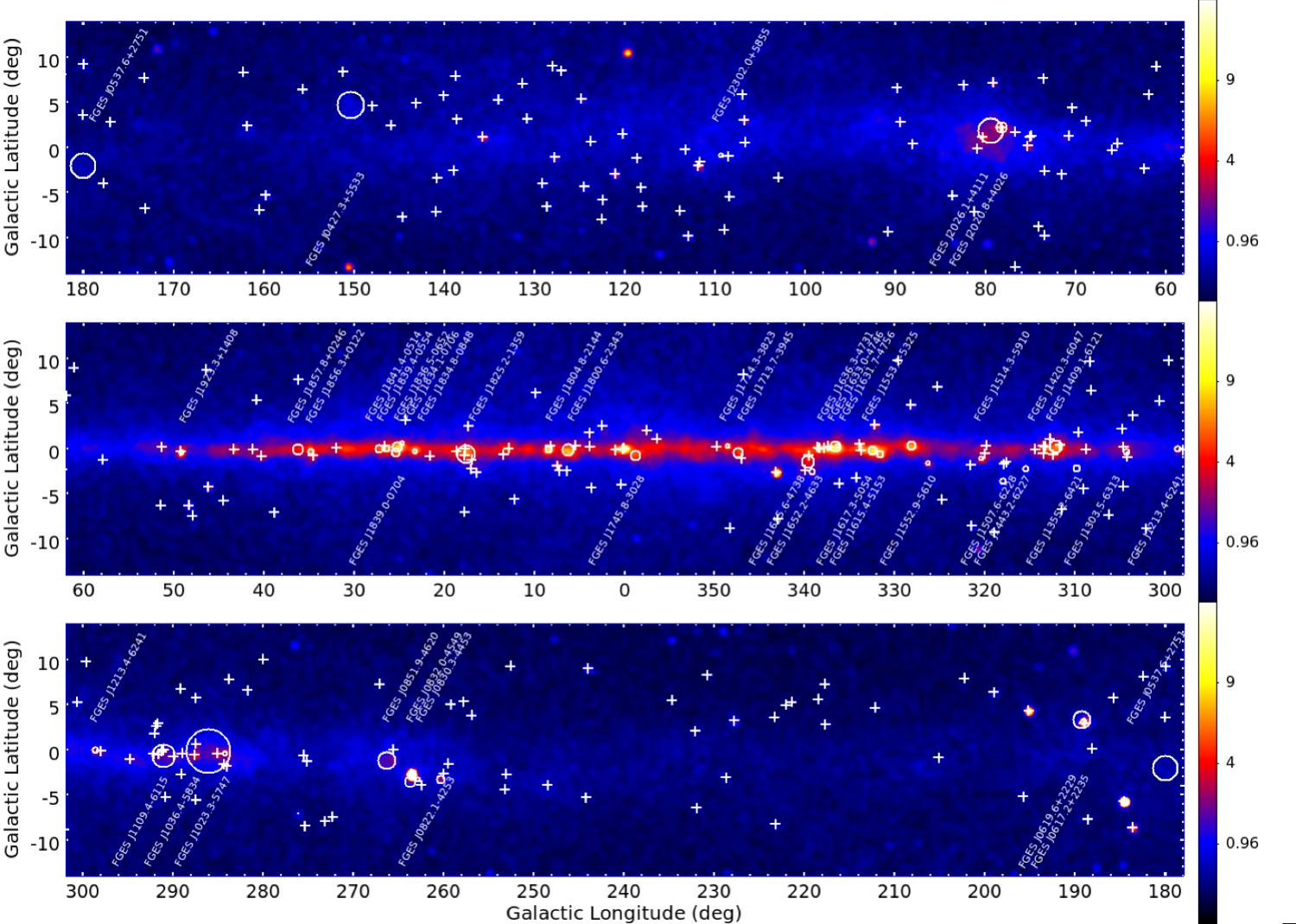}
\caption{Smoothed count map in the 10\,GeV~--~2\,TeV band represented in Galactic coordinates and cartesian projection. The image has been smoothed with a Gaussian kernel with a size of 0.5$^{\circ}$. The color scale is square root and the units are counts per (0.1$^{\circ}$)$^2$. White circles indicate the position and extension of the 46 extended sources described in this work. White crosses mark the location of point sources.
    \label{fig:cmap}}
\end{centering}
\end{figure*}

We used 6 years (from 2008 August 4 to 2014 August 4) of Pass~8 SOURCE photons with reconstructed energy in the 10 GeV~--~2 TeV range. Photons detected at zenith angles larger than 105$^\circ$ were excised to limit the contamination from $\gamma$-rays generated by cosmic-ray interactions in the upper layers of the atmosphere. Moreover, data were filtered removing time periods when the instrument was not in sky-survey mode. \emph{Fermi} Science Tools v10r01p01 and instrument response functions (IRFs) P8R2\_SOURCE\_V6 were used for this analysis. In addition the analysis was restricted to regions within $7^{\circ}$ from the Galactic plane.
Figure~\ref{fig:cmap} shows a count map of the Galactic plane observed by the \emph{Fermi}-LAT above 10 GeV highlighting large structures with a Gaussian smoothing radius of 0.5$^{\circ}$. The bright remnants IC~443 (l=189.06$^{\circ}$) and $\gamma$ Cygni (l=78.15$^{\circ}$) stand out clearly, but a large number of other sources are also apparent. Several are coincident with higher-energy sources detected by ground-based $\gamma$-ray experiments, such as the Kookaburra complex (l=313.38$^{\circ}$), and will be discussed in Section~\ref{section:results}. The large number of sources visible in the map highlights the excellent sensitivity and angular resolution of the \emph{Fermi}-LAT at high energies afforded by the new Pass 8 data.

\section{Detection of new extended sources}
\label{subsection:Analysis}
\subsection{Input source model construction}\label{subsection:ROI}
The analysis of the full data set was divided into smaller regions of the sky each of which must be represented by a spectral and spatial model. For each region, we start with a sky model that includes all point-like and extended \emph{Fermi}-LAT sources listed in the 3FGL catalog, the Galactic diffuse and isotropic emission, and pulsars from the Second \emph{Fermi} LAT Pulsar Catalog \citep{2013ApJS..208...17A} as well as from 3FGL. The energy range used in this work prevents a reasonable fit of a pulsar spectral component modeled by a power law with an exponential cutoff. The bulk of the pulsar emission and the cutoff energy of the pulsar, typically a few GeV, lie below 10\,GeV. Therefore, we decided to fit only the normalization and index for pulsars while keeping the cutoff energy as a fixed parameter.
The Galactic diffuse emission was modeled by the standard \emph{Fermi}-LAT diffuse emission ring-hybrid model gll\_iem\_v06.fits~\citep{2016ApJS..223...26A}, and the residual background and extragalactic radiation were described by a single isotropic component with the spectral shape in the tabulated model iso\_P8R2\_SOURCE\_V6\_v06.txt. The models are available from the \emph{Fermi} Science Support Center (FSSC)\footnote{\url{http://fermi.gsfc.nasa.gov/ssc/}}. In the following, we fit the normalizations of the Galactic diffuse and the isotropic components.

\subsection{Analysis method}\label{subsection:fit}
Two different software packages for maximum-likelihood fitting were
used to analyze \emph{Fermi}-LAT data: $\mathtt{pointlike}$ and $\mathtt{gtlike}$. These tools fit \emph{Fermi}-LAT data with a parametrized model of the sky, including models for the instrumental, extragalactic and Galactic components of the background. 
$\mathtt{pointlike}$ is a software package \citep{2010PhDT.......147K} validated by \cite{2012ApJ...756....5L} that we used to fit the positions of point-like sources in the region of interest (ROI) and fit the spatial parameters of spatially-extended sources presented in Section~\ref{subsection:SourceDetection}. $\mathtt{gtlike}$ is the standard maximum-likelihood method distributed in the Fermi Science Tools by the FSSC. We apply it in binned mode, combining the four P8R2\_SOURCE\_V6 PSF event types in a joint likelihood function.

In the following analysis, we used $\mathtt{pointlike}$ to evaluate the best-fit position and extension, as well as preliminary spectral values, for each new source added in our model. Using those morphologies, we subsequently employed $\mathtt{gtlike}$ to obtain the best-fit spectral parameters (initializing spectra at the $\mathtt{pointlike}$-determined values) and statistical significances (see Section~\ref{subsection:SourceSpectra}). Both methods agree with each other for all derived quantities, but all spectral parameters and significances quoted in the text were obtained using $\mathtt{gtlike}$.

Since the $\mathtt{pointlike}$ and $\mathtt{gtlike}$ analyses use circular and square ROI geometries, respectively, we included photons within a radius of $10^{\circ}$ when using $\mathtt{pointlike}$ to characterize the whole ROI and within a $10^{\circ} \times 10^{\circ}$ square region centered on the extended source of interest when using $\mathtt{gtlike}$ to perform the spectral analysis. Both analyses use an energy binning of 8 bins per decade and the MINUIT\footnote{For more information about MINUIT see \url{http://lcgapp.cern.ch/project/cls/work-packages/mathlibs/minuit/doc/doc.html}} optimizer for likelihood fitting.

\begin{figure*}[!ht]
\begin{centering} 
\includegraphics[width=0.9\textwidth]{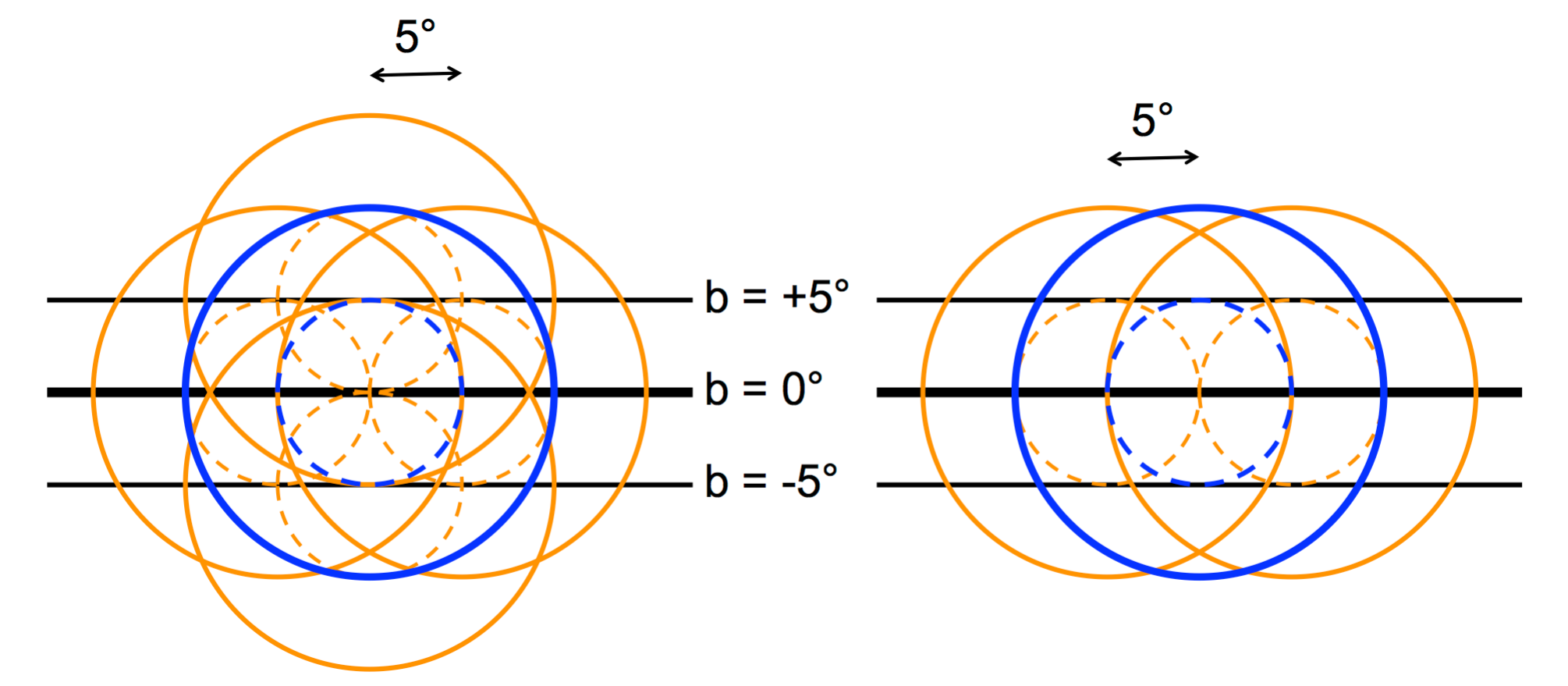}
\caption{Schematic representations of the arrangement of the analysis regions for the two pipelines used for localization and extension. \textit{Left:} description of the main pipeline, defined in Section~\ref{subsection:SourceDetection}; each ROI (solid blue circle) of radius 10$^\circ$ is centered on $b = 0^\circ$ and $\pm$~5$^\circ$ and separated from its neighboring ROIs (orange circles) by 5$^\circ$ in Galactic longitude; all sources within 5$^\circ$ of the center (dashed circle with the same color) were set free for the primary pipeline. \textit{Right:} details of the secondary pipeline described in Appendix~\ref{appen:xCheck}: each ROI (marked by a solid blue circle) of radius 10$^\circ$ is centered on $b = 0^\circ$ and separated from its neighbors (orange circles) by 5$^\circ$ in Galactic longitude.} \label{fig:pipeline}
\end{centering}
\end{figure*}

\subsection{Localization and extension}\label{subsection:SourceDetection}
We developed an analysis pipeline, similar to that used in the 2FHL catalog. We describe here this pipeline and present an alternate analysis implemented as a cross-check in Appendix \ref{appen:xCheck}. Our pipeline was launched over 216 ROIs of radius 10$^\circ$, centered on b = 0$^\circ$ and $\pm$~$5^\circ$ with overlapping neighboring ROIs separated by 5.0$^\circ$ in Galactic longitude (Figure~\ref{fig:pipeline}). To homogenize the analysis, extended sources were all fit assuming a uniform disk shape. The pipeline included extended sources from the 3FGL catalog, which were initialized at their best-fit disk extension. If the source was previously modeled with a Gaussian shape in the 3FGL catalog, we initialized the disk radius at $1.85~\sigma_{Gaussian}$ as suggested by \cite{2012ApJ...756....5L}. If the source was previously modeled with a multi-wavelength template, we used the average between the semi-major and semi-minor axes, reported in the 3FGL catalog, to initialize the disk radius.
In each region the procedure to find all point and extended sources proceeded as follows using $\mathtt{pointlike}$.
\begin{enumerate}
\item Using the initial sky model defined above, the first step of our pipeline aims to find the best spectral parameters for all free sources in the region using $\mathtt{pointlike}$. All sources within 5$^\circ$ of the center were set free. The significance of each source was evaluated using the test statistic ${\rm TS}=2(\ln \mathcal{L}_1 - \ln \mathcal{L}_0)$, where $\mathcal{L}_0$ and $\mathcal{L}_1$ are the likelihoods of the background (null hypothesis) and the hypothesis being tested (source plus background). The formal statistical significance of this test can be obtained from Wilks' theorem \citep{STMAZ.03029575}. In the null hypothesis, TS follows a $\chi^2$ distribution with n degrees of freedom where n is the number of additional parameters in the model. At each step in the procedure, sources with TS$<$16 were removed from the model.
\item Due to the 6 years integration time of our analysis compared to the 4 years of the 3FGL catalog, we expected to find new statistically-significant sources (TS $>$ 16). To detect these new sources, we generated a TS map for a point source with a Power-Law spectral index of 2.0 including all significant 3FGL sources in the background model. The TS map covered $7^\circ \times 7^\circ$ length with $0.1^{\circ}$ pixels. We added a source at the location of every peak with TS above 16 which was separated by more than 0.2$^\circ$ from another peak in the TS map (or source in the region) and then fit them iteratively for extension starting from the brightest one. {This means that all extended sources detected by our pipeline must be first detected as a point source with a TS higher than 16. This is a limit of the method employed here and we can expect that very extended sources where the surface brightness is too faint will not be detected here.} If the TS of an added source became smaller than 16 during the iterative process, the source was removed and the localization, extension and spectrum of all sources located within 0.5$^\circ$ were refit (including the localization of 3FGL sources). The threshold to define a source as extended is set as $\rm TS_{ext} \ge 16$, where ${\rm TS_{ext} = 2~ln(\mathcal{L}_{ext} / \mathcal{L}_{ps})}$ \citep{2012ApJ...756....5L}, i.e twice the logarithm of the likelihood ratio of an extended to a point source. The choice of a threshold $\rm TS_{ext}$ set to 16 corresponds to a formal $4\sigma$ significance\footnote{Using 20000 statistically independent simulations, \cite{2012ApJ...756....5L} showed that the cumulative density of $\rm TS_{ext}$ follows a $\chi^2$ distribution with one degree of freedom.}. If this threshold was met then the disk-modeled source was kept in the ROI. We stopped adding sources when the source TS was less than 16.
\item Again, due to the different integration time and energy range, we might see variations in morphology for already detected extended sources. The spatial and spectral parameters of all 3FGL sources are therefore refit once these new point sources and extended sources are added in the source model of each region.
\item As a last step, to address the ambiguity between detecting a source as spatially extended as opposed to a combination of point sources, we utilized the algorithm detailed in \cite{2012ApJ...756....5L} to simultaneously fit the spectra and positions of two nearby point sources. To help with convergence, it begins by dividing the extended source into two spatially coincident point-like sources and then fitting the sum and difference of the positions of the two sources without any limitations on the fit parameters. We only considered a source to be extended if TS$\rm{_{ext}}$ $>$ TS$\rm{_{2pts}}$ (improvement when adding a second point source defined as ${\rm TS_{2pts} = 2~ln(\mathcal{L}_{2pts} / \mathcal{L}_{ps})}$). If an extended source did not meet this criterion, it was then replaced by two point sources located at the best positions found by the above algorithm. It should be noted that ${\rm TS_{2pts}}$ cannot be quantitatively compared to TS$\rm{_{ext}}$ using a simple likelihood-ratio test to evaluate which model is preferred because the models are not nested. As an alternative, we can consider the Akaike information criterion test \citep[AIC,][]{1974ITAC...19..716A}. The \aic is defined as AIC = $2k - 2~\rm{ln}\mathcal{L}$, where $k$ is the number of parameters in the model. In this formulation, the best hypothesis is considered to be the one that minimizes the \aic. The two point-like sources hypothesis has three more parameters than the single extended source hypothesis (two more spatial parameters and two more spectral parameters compared to one extension parameter), so the comparison AIC$\rm{_{ext}} <$ AIC$\rm{_{2pts}}$ is formally equivalent to TS$\rm{_{ext}} + 6$ $>$ TS$\rm{_{2pts}}$. This means that our criterion is more restrictive than the AIC test. It was extensively tested in \cite{2012ApJ...756....5L} using simulations showing that TS$\rm{_{ext}}$ $>$ TS$\rm{_{2pts}}$ is a powerful test to avoid cases of simple confusion of two point-like sources. But it could always be the case that an extended source is actually the superposition of multiple point-like or extended sources that could be resolved with deeper observations of the region.
\item When the sky model was complete, all new sources were tested for spectral curvature using a log-normal model (referred to as LogParabola or LogP with a curvature noted $\beta$). We assessed the significance of the spectral curvature for a given source by ${\rm TS_{\rm curve} = 2~ln(\mathcal{L}_{LogP} / \mathcal{L}_{PL})}$. Since the Power-Law is a special case of LogParabola (with $\beta$ = 0) and $\beta$ = 0 is inside the allowed interval, we expect that TS$_{\rm curve}$ is distributed as $\chi^2$ with one degree of freedom. We switched to LogParabola and refit the ROI if TS$_{\rm curve} > 16$, corresponding to 4$\sigma$ significance for the curvature. Only one extended source shows such curvature.
\item To complete the construction of the source model of the region, we take the output of the previous steps for the 4 surrounding ROIs plus the ROI of interest as defined in color in the left panel of Figure~\ref{fig:pipeline} using a $10^{\circ}$ radius centered on a Galactic latitude of b = 0$^\circ$. Sources appearing in multiple ROIs are defined using the parameters obtained in the closest ROI center. We refit the spatial parameters of any previously added extended sources within 5$^\circ$ of the center (starting from the highest TS value) as well as the spectra of sources in this region, while all other sources in the ROI were fixed. This allows a direct comparison of the two pipelines since the size, location and free radius of the regions are then identical. 
\end{enumerate}

This analysis detected 51 sources with TS$_{\rm ext}$ $>$ 16, TS $>$ 25 and TS$_{\rm ext}$ $>$ TS$_{\rm 2pts}$. Spectral and spatial parameters for the detected extended sources are compatible in both the analysis described above, and the secondary pipeline described in Appendix \ref{appen:xCheck}, in most cases. Only 2 detected sources were rejected: one undetected by the main pipeline and another one undetected by the secondary pipeline. They are discussed in the Appendix. The morphological results derived by $\mathtt{pointlike}$ for the 46 sources which also pass the same TS criteria with $\mathtt{gtlike}$ (see below) are presented in Table~\ref{tab:morpho}. It should be noted that the final list of point sources detected by our two pipelines agree perfectly with those reported by the 3FHL catalog in the latitude range $\pm \, 7^{\circ}$ \citep{3FHL} using 27 spatial templates derived in this analysis (either when the extended source is newly detected here or when the model provides a better representation of the source). The point sources not detected by our pipeline (less than 10\%) are all low TS sources (close to our threshold of 25) and can be explained by the reduced dataset and binned analysis used here.

\subsection{Spectra}\label{subsection:SourceSpectra}
The \emph{Fermi}-LAT spectra of the detected extended sources were derived by $\mathtt{gtlike}$ assuming the best uniform disk extension found by $\mathtt{pointlike}$ in Section~\ref{subsection:SourceDetection}. The $\mathtt{gtlike}$ analysis was used to fit the spectral parameters of each source but also its associated TS, TS$_{\rm ext}$, TS$_{\rm 2pts}$ and TS$_{\rm curve}$. Since $\mathtt{gtlike}$ makes fewer approximations in calculating the likelihood, spectral parameters found with $\mathtt{gtlike}$ are slightly more accurate and this cross-check is extremely useful. Only three sources were rejected at this step because they did not meet the threshold in terms of TS, TS$_{\rm ext}$ or TS$_{\rm 2pts}$: 
\begin{itemize}
\item The Crab Nebula which is detected with a TS$_{\rm ext}$ of 30 for an extension of 0.03$^{\circ}$ with $\mathtt{pointlike}$ and 0 with $\mathtt{gtlike}$. This discrepancy can be explained by the complexity of fitting the nebula simultaneously to its associated pulsar. In this case, the extension found by $\mathtt{pointlike}$ is not preferred by $\mathtt{gtlike}$ over a simple point source.
\item HESS~J1640$-$465 which is detected with a TS$_{\rm ext}$ of 18 for an extension of $0.08^{\circ} \pm 0.02^{\circ}$ with $\mathtt{pointlike}$ and only 10 with $\mathtt{gtlike}$; \cite{2014ApJ...794L..16L} reported a Gaussian size for this source of $0.07^{\circ}$ (with a TS$_{\rm ext}$ value of only 6) equivalent to a disk size of $0.13^{\circ}$ above 3 GeV. This radius is larger than the disk size reported here using $\mathtt{pointlike}$, and may explain the low TS value obtained in our $\mathtt{gtlike}$ analysis since we fixed the extension value obtained with $\mathtt{pointlike}$.
\item An unidentified source detected at (l, b) = (292.05$^\circ$, 2.66$^\circ$) for which the $\mathtt{gtlike}$ calculated TS$_{\rm 2pts}$ is  greater than TS$_{\rm ext}$.
\end{itemize}
All TS values for the remaining 46 sources are presented in Table~\ref{tab:morpho} while their spectral parameters are listed in Table~\ref{tab:spectra}.
In addition to performing a spectral fit over the entire energy range, we computed an SED by fitting the flux of the source independently in 4 energy bins spaced uniformly in log space from 10 GeV to 2 TeV. During this fit, we fixed the spectral index of the source at 2 as well as the model of background sources to the best fit obtained in the whole energy range except the Galactic diffuse background and the prefactor of sources closer than 5$^{\circ}$. We defined a detection in an energy bin when TS $\ge 4$ and otherwise computed a 95\% confidence level flux upper limit. The upper limit is obtained by looking for 2$\Delta$ln(likelihood) = 4 when increasing the flux from the maximum likelihood value if the TS value of the source is larger than 1. Whenever TS $<$ 1 we switched to the Bayesian method proposed by \cite{Helene83}.    

\subsection{Systematic errors}\label{subsection:systematics}
Three main systematic uncertainties can affect the extension fit and the spectra of the detected extended sources: uncertainties in our model of the Galactic diffuse emission, uncertainties on the shape of the extended source and uncertainties in our knowledge of the \emph{Fermi}-LAT IRFs. This last contribution was estimated using custom IRFs chosen to maximize and minimize effective area and PSF within their systematic uncertainty bands\footnote{\url{https://fermi.gsfc.nasa.gov/ssc/data/analysis/LAT_caveats.html}. The uncertainty in the IRFs does not affect the spectra by more than 5\% and can be safely neglected in this study.} Then, to explore the systematic effects on our sources' fitted properties caused by interstellar emission modeling, we have followed the prescription developed in \cite{2016ApJS..224....8A}. Each extended source was refit using 8 alternate interstellar emission models (IEMs) and, for each fitted parameter $P$ (namely the disk extension, the integrated flux above 10 GeV and the spectral index), we obtained a set of 8 values $P_i$ that we compared to the value obtained with the standard model $P_\mathrm{STD}$ following Equation (5) in \cite{2016ApJS..224....8A}. The corresponding systematic error for each source and for these 3 parameters is reported in Tables~\ref{tab:morpho} and \ref{tab:spectra}. We encountered convergence issues when fitting the extension of three sources with a fraction of the 8 alternate diffuse models: the source at the Galactic Center FGES J1745.8$-$3028, the Cygnus cocoon FGES J2026.1+4111, and FGES J0832.0$-$4549 in the region of Vela-X. The number of alternate diffuse models used is written in parentheses in column 7 of Table~\ref{tab:morpho} for these three cases. Finally, as noted above, the imperfect knowledge of the true $\gamma$-ray morphology introduces a last source of error. To provide a feeling of the influence of the assumed source shape, we refitted all sources using a 2D-Gaussian model. This spatial model does not offer a good representation for shell-type SNRs such as RX~J1713.7$-$3946 but is well adapted to PWNe-type sources for which the $\gamma$-ray signal is expected to be visible up to large distances. Table~\ref{tab:gauss} gives the morphological and spectral parameters of this Gaussian fit. Please note that all errors are statistical only since this table is only provided as a cross-check. It is clear from this table that the majority of the extended sources are very stable with respect to the assumed shape except confused sources and/or very large sources.

\section{Discussion}\label{section:results}
We detected 46 statistically-significant spatially-extended \emph{Fermi}-LAT $\gamma$-ray sources as well as 162 point-like sources in the $\pm 7^{\circ}$ latitude range as can be seen in Figure~\ref{fig:cmap}. The results of the spatial and spectral analyses for the extended sources are shown in Tables~\ref{tab:morpho} and \ref{tab:spectra}. Among these extended sources, 16 are new, 13 are in agreement with previous publications and 17 have a different morphology (we defined the criterion for significant difference with respect to previously published values as $\Delta_{\rm{FGES - Published}} > 2 \sqrt{(\sigma_{\rm{FGES}}^2 + \sigma_{\rm{Published}}^2)}$, $\sigma$ being the uncertainty on the parameter of interest). In the latitude interval covered by our search, only four Galactic sources already detected as significantly extended in previous works are not detected in this work: HB21, HB3, HB9 and W3. These four sources are also not detected in the 3FHL catalog \citep{3FHL} using their associated morphological templates. 


\subsection{Agreement with previous publications}
\label{subsection:agree}
\begin{figure*}[ht]
\begin{center}
\begin{tabular}{ll}
\includegraphics[width=0.98\textwidth]{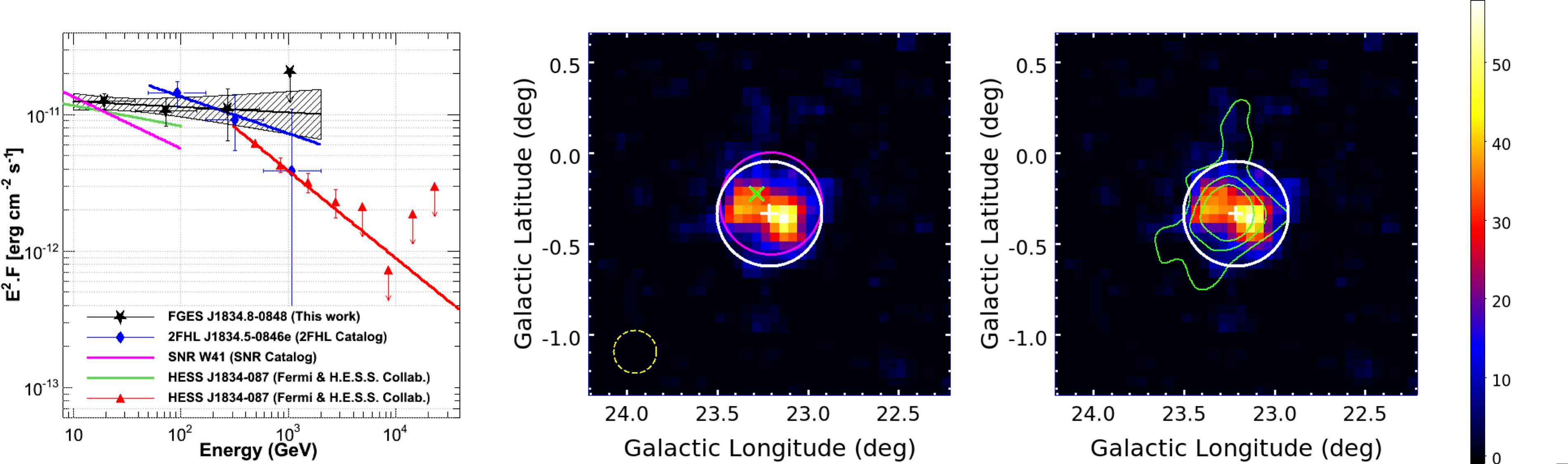}\\
\includegraphics[width=0.98\textwidth]{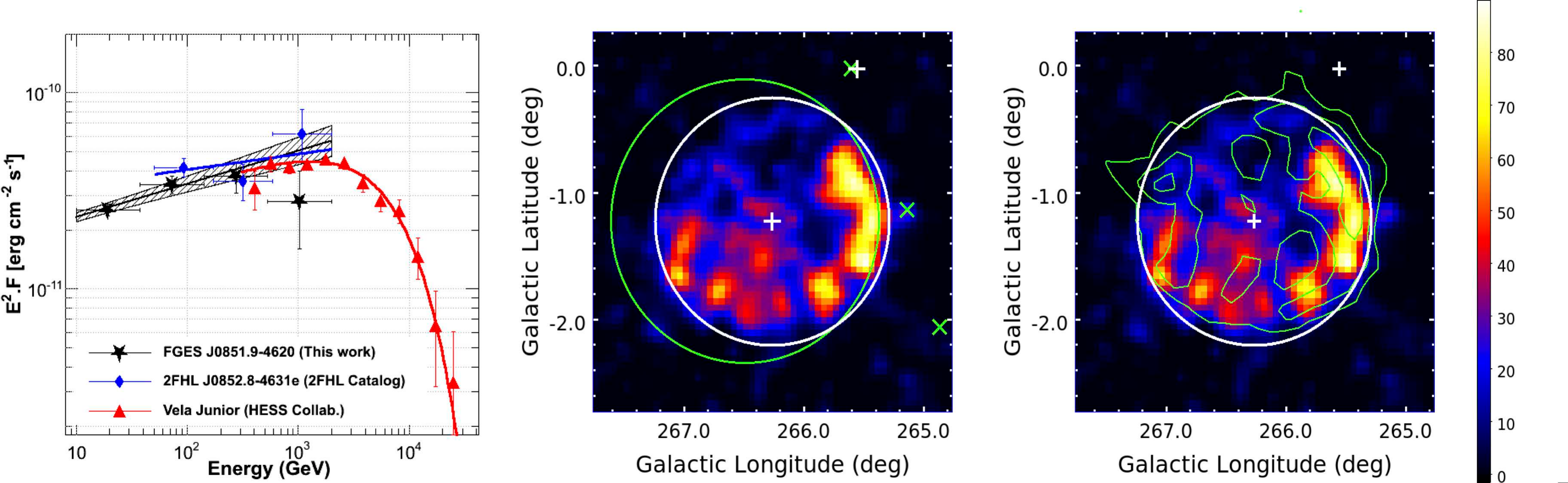}\\
\end{tabular}
\end{center}
\caption{
\label{fig:agree1}Spectral (\emph{left}) and spatial (\emph{middle} and \emph{right}) properties of the extended sources associated with W41 (top) and the SNR Vela Junior (bottom). \emph{Left:} Spectral energy distributions with data points from this analysis (black stars and dashed butterfly), from the SNR catalog \cite[purple line,][]{2016ApJS..224....8A}, 2FHL catalog (blue diamonds and line), previous \emph{Fermi}-LAT publication \citep[green line, ][]{2015A&A...574A..27H} and IACT data \citep[red triangles and line,][ for W41 and Vela Junior respectively]{2015A&A...574A..27H, 2007ApJ...661..236A}. \emph{Middle:} Background-subtracted TS map with the Galactic diffuse and isotropic emission and surrounding point sources included in the model to highlight the location of emission coming from the extended source. White circles and central crosses indicate the disk extension and centroid as fit in this work. Green and purple markings present the position of point-like and extended sources published in the 3FGL and 2FHL catalogs respectively. The yellow dashed circle in the bottom left corner of the top Figure illustrates the PSF size of the instrument for the analysis carried in this article. \emph{Right:} Same TS map, but with IACT contours (green, from the above-quoted references) overlaid.
}
\end{figure*}

\begin{figure*}[ht]
\begin{center}
\begin{tabular}{ll}
\includegraphics[width=0.98\textwidth]{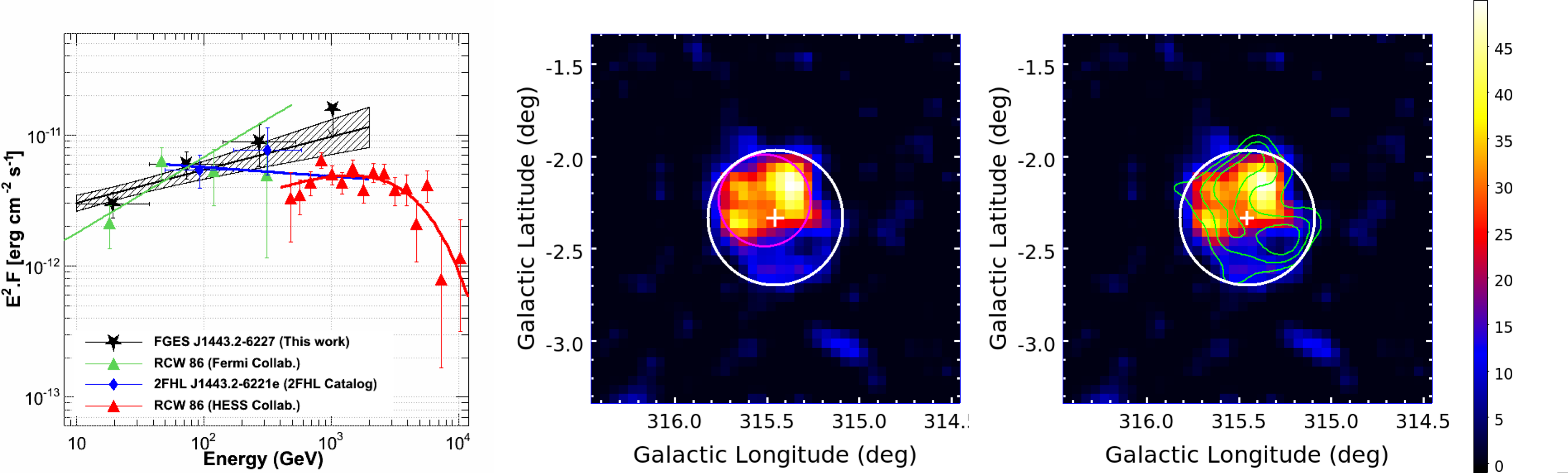}\\
\includegraphics[width=0.98\textwidth]{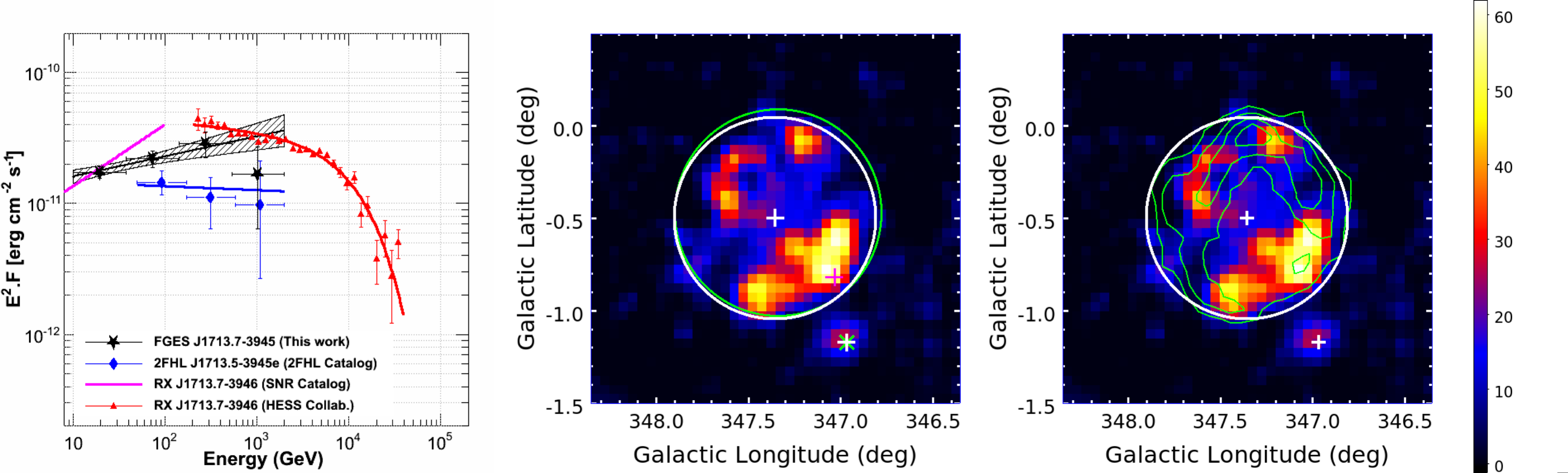}\\
\end{tabular}
\end{center}
\caption{
\label{fig:agree2}Extended sources associated with the SNR RCW 86 (top) and RX~J1713.7$-$3946 (bottom) following conventions of Figure~\ref{fig:agree1} and identical references for the 2FHL and SNR catalogs. Left: Spectral energy distributions of the extended sources with data points from this analysis (black stars and dashed butterfly), from the SNR catalog (purple line), 2FHL catalog (blue diamonds and line), previous \emph{Fermi}-LAT publication \cite[green triangles and line,][]{2016ApJ...819...98A} and IACT data \citep[red triangles and line,][ for RCW 86 and RX~J1713.7$-$3946 respectively]{2009ApJ...692.1500A, 2007A&A...464..235A}. Middle and right: Background-subtracted TS maps using the same conventions of Figure~\ref{fig:agree1} and above-quoted references for the IACT contours shown in green. White circles and central crosses indicate the disk extension and centroid as fit in this work. In the upper middle panel, the grey circle corresponds to the extension found in \cite{2016ApJ...819...98A}. 
}
\end{figure*}

\begin{figure*}[ht]
\begin{center}
\begin{tabular}{ll}
\includegraphics[width=0.98\textwidth]{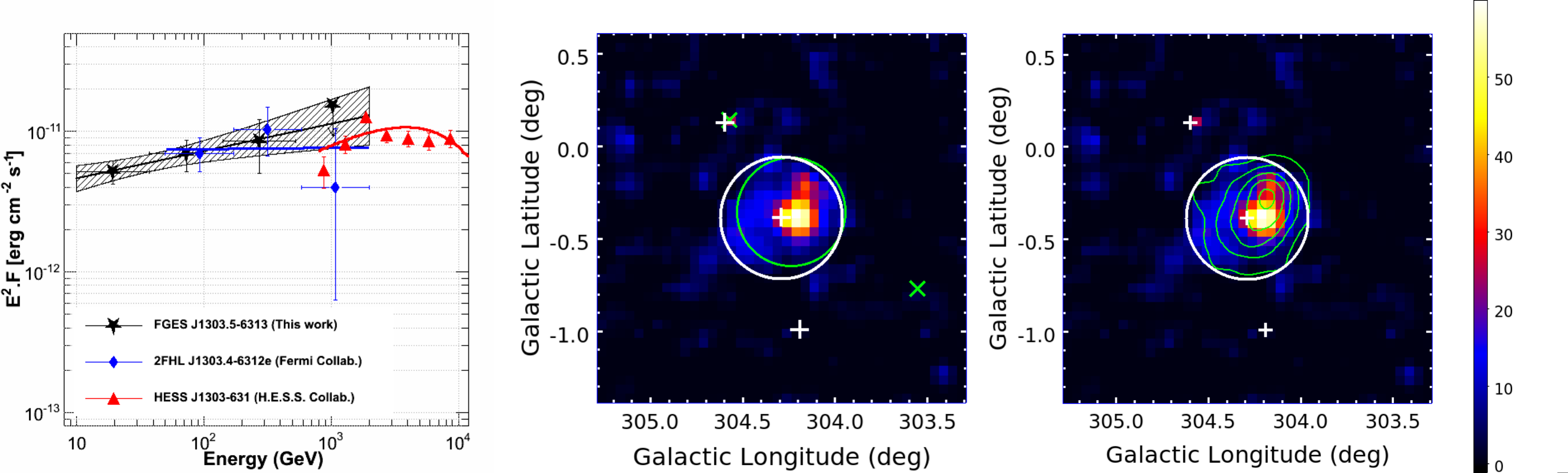}\\
\includegraphics[width=0.98\textwidth]{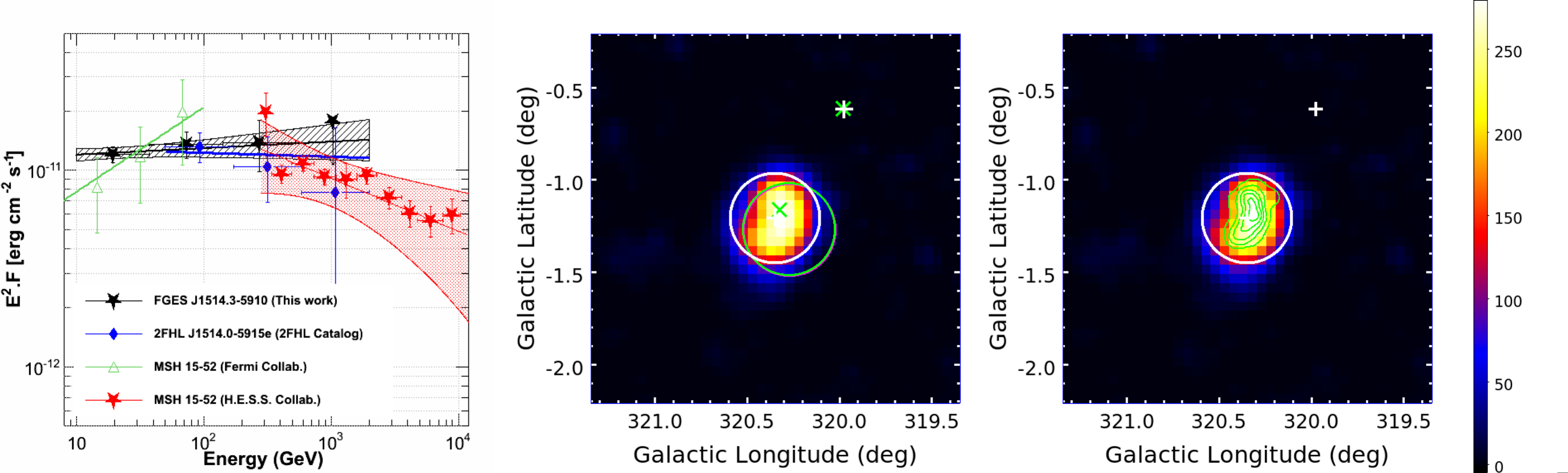}\\
\end{tabular}
\end{center}
\caption{
\label{fig:agree3}Extended sources associated with the PWNe HESS~J1303$-$631 (top) and MSH 15$-$52 (bottom) following conventions of Figure~\ref{fig:agree1}. Left: Spectral energy distributions of the extended sources with data points from this analysis (black stars and dashed butterfly), 2FHL catalog (blue diamonds) and IACT data \citep[red line,][ for HESS J1303$-$631 and MSH 15$-$52 respectively]{2012A&A...548A..46H, 2005A&A...435L..17A}. Middle and right: Background-subtracted TS maps using the same conventions of Figure~\ref{fig:agree1} and above-quoted references for the IACT contours shown in green. White circles and central crosses indicate the disk extension and centroid as fit in this work.
}
\end{figure*}

\begin{figure*}[ht]
\begin{center}
\begin{tabular}{ll}
\includegraphics[width=0.98\textwidth]{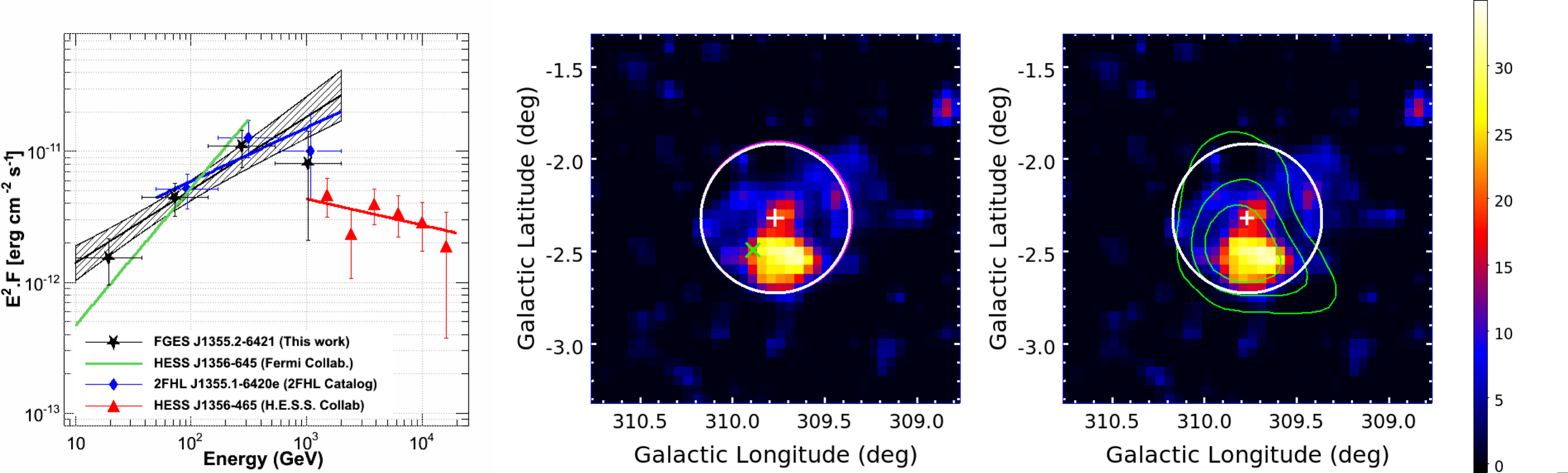}\\
\includegraphics[width=0.98\textwidth]{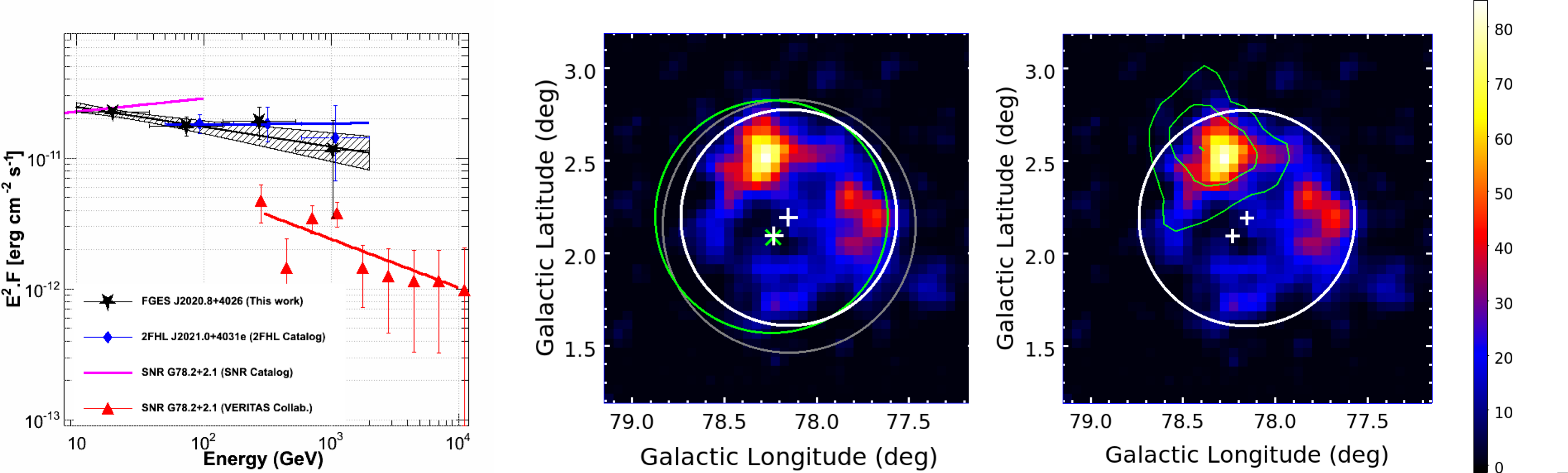}\\
\end{tabular}
\end{center}
\caption{
\label{fig:agree4}Extended sources associated with the PWN HESS~J1356$-$645 (top) and the SNR $\gamma$ Cygni (bottom) following conventions of Figure~\ref{fig:agree1}. Left: Spectral energy distributions with data points from this analysis (black stars and dashed butterfly), 2FHL catalog (blue diamonds and line), previous \emph{Fermi}-LAT publication \cite[in green,][]{2013ApJ...773...77A} and IACT data \citep[red triangles and line,][ for HESS~J1356$-$645 and $\gamma$ Cygni respectively]{2011A&A...533A.103H, 2013ApJ...770...93A}. Middle and right: Background-subtracted TS maps using the same conventions of Figure~\ref{fig:agree1} and above-quoted references for the IACT contours shown in green, and SNR catalog disk size in grey (bottom). White circles and central crosses indicate the disk extension and centroid as fit in this work.
}
\end{figure*}

The 13 sources in agreement with previous publications are:
\begin{itemize}
\item FGES J0617.2+2235 (associated with the SNR IC~443), 
\item FGES J0851.9$-$4620 (associated with the SNR Vela Junior), 
\item FGES J0822.1$-$4253 (associated with Puppis~A), 
\item FGES J1303.5$-$6313 (associated with HESS J1303$-$631), 
\item FGES J1355.1$-$6420 (associated with the PWN HESS J1356$-$645~\footnote{A typo was recently discovered in the disk extension value reported in Table 5 of \cite{2016ApJS..222....5A} and in its associated fits file. An erratum is being prepared quoting a value of 0.41$^\circ$ for this source.)}, 
\item FGES J1443.2$-$6227 (associated with the SNR RCW~86), 
\item FGES J1514.3$-$5910 (associated with MSH 15$-$52), 
\item FGES J1552.9$-$5610 (associated with MSH 15$-$56), 
\item FGES J1615.4$-$5153 (associated with HESS~J1614$-$518), 
\item FGES J1713.7$-$3945 (associated with the SNR RX~J1713.7$-$3946), 
\item FGES J1834.8$-$0848 (associated with W41), 
\item FGES J1834.1$-$0706 (associated with the SNR G24.7+0.6), 
\item FGES J2020.8+4026 (associated with $\gamma$ Cygni). 
\end{itemize}
Figures~\ref{fig:agree1}, \ref{fig:agree2}, \ref{fig:agree3} and \ref{fig:agree4} (top) provide the background-subtracted TS maps (i.e. TS maps with all components other than the source included in the model) and SEDs for eight which are detected at TeV energies, showing an excellent agreement with the results obtained by the H.E.S.S. experiment. The complete shells of RX~J1713.7$-$3946 and Vela Junior appear in the background-subtracted TS maps while RCW~86 presents a brighter emission on the northern part of the remnant where fast shocks and a low density medium has been measured by \cite{2006ApJ...648L..33V}, \cite{2009Sci...325..719H} and \cite{2008PASJ...60S.123Y}. The GeV extension of the PWN HESS~J1303$-$631 seems to be in slight disagreement with the previously published value, however it is consistent within the large uncertainties of $0.09^{\circ}_{\rm stat} \pm 0.10^{\circ}_{\rm syst}$ derived at that time with only 45 months of data. The region including FGES J1834.1$-$0706 (close to the H.E.S.S. source HESS J1837$-$069) is described in section~\ref{subsection:confusion} while the region of the SNR IC~443 and its surrounding is discussed in section~\ref{subsection:new}. 

\begin{itemize}
\item {\bf The SNR $\gamma$ Cygni (FGES~J2020.8+4026):}
$\gamma$ Cygni (SNR G78.2+2.1) is a nearby ($\sim$1.7 kpc) middle-aged SNR already detected by \emph{Fermi}-LAT in different energy bands \citep{2016ApJS..222....5A, 2016ApJS..224....8A}. Our analysis is in perfect agreement with previous publications of the SNR. It still shows a much higher flux in comparison to the TeV signal detected by VERITAS from VER J2019+407 above 300 GeV, as can be seen in Figure~\ref{fig:agree4} (bottom left). The TeV signal is more compact (Figure~\ref{fig:agree3} middle and right) and coincides with the brightest part of the northern radio shell, opposite to molecular material locations~\citep{2013ApJ...770...93A}. VER J2019+407's nature and relationship to the emission detected by \emph{Fermi}-LAT thus remains unclear and extremely puzzling since VERITAS should in principle see emission from the majority of the SNR according to the new spectrum derived in this analysis for a uniform disk encompassing the whole shell, as already stated by \citet{2015EPJWC.10504005W}. Interestingly, a recent publication by \cite{2016ApJ...826...31F} shows that the \emph{Fermi}-LAT spectrum on VER J2019+407 alone is harder than the rest of the shell, with indices 1.8 below a break energy of 71 GeV and 2.5 above the break. A detailed spectrally-resolved morphological analysis of the \emph{Fermi}-LAT emission is required to better constrain the model parameters and the nature of the radiation.

\end{itemize}

\subsection{Differences with previous publications}
Differences between this work and previous publications can be explained in four ways: use of a different morphological template to model the extended source, use of a different energy threshold, improvements in analysis methods and/or increased statistics, and ambiguities from source confusion and contamination.

\subsubsection{Effect of the morphological template}
\begin{figure*}[ht]
\begin{center}
\begin{tabular}{ll}
\includegraphics[width=0.98\textwidth]{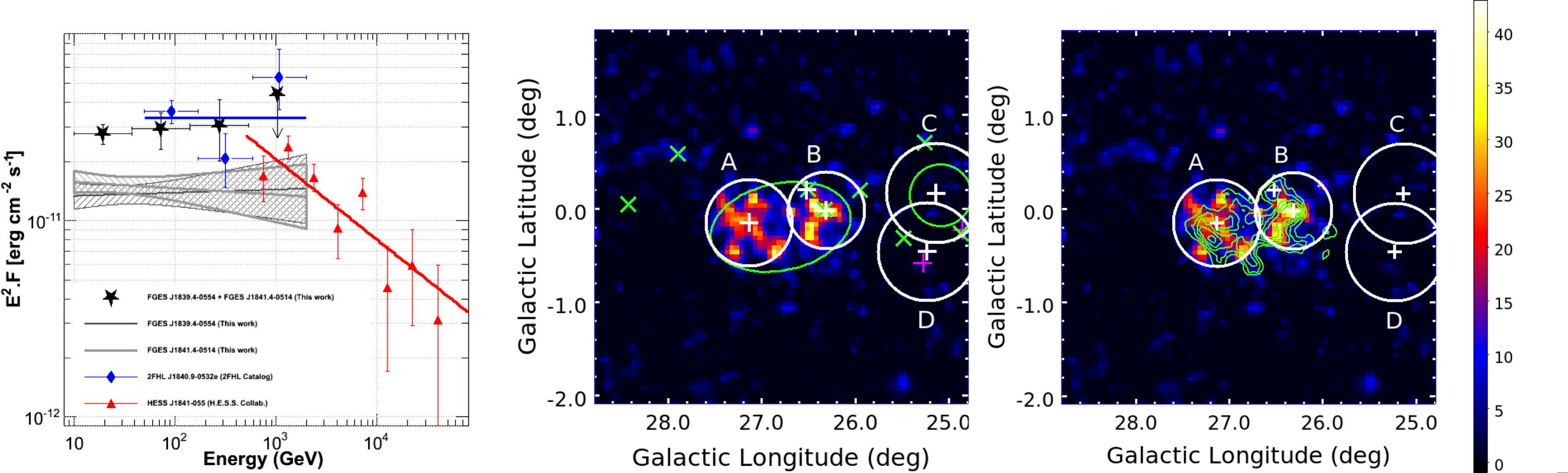}
\end{tabular}
\end{center}
\caption{
\label{fig:hessj1841}Left: Spectral energy distributions of the unidentified source HESS~J1841$-$055 with data points from this analysis (black and grey dashed butterflies for FGES J1839.4$-$0554 and FGES J1841.4$-$0514 respectively), 2FHL catalog (blue diamonds and line) and IACT data \cite[red line, ][]{2008A&A...477..353A}. The black stars represent the sum of the emission of the two coincident extended sources obtained from this analysis. Middle and right: Background-subtracted TS maps of HESS~J1841$-$055 using the same conventions of Figure~\ref{fig:agree1} and above-quoted reference for the IACT contours shown in green. White circles and central crosses indicate the disk extension and centroid as fit in this work. The letters A, B, C, D indicate the FGES sources FGES J1841.4$-$0514, FGES J1839.4$-$0554, FGES~J1836.5$-$0652 and FGES J1839.0$-$0704 respectively.
}
\end{figure*}

\begin{figure*}[ht]
\begin{center}
\begin{tabular}{ll}
\includegraphics[width=0.98\textwidth]{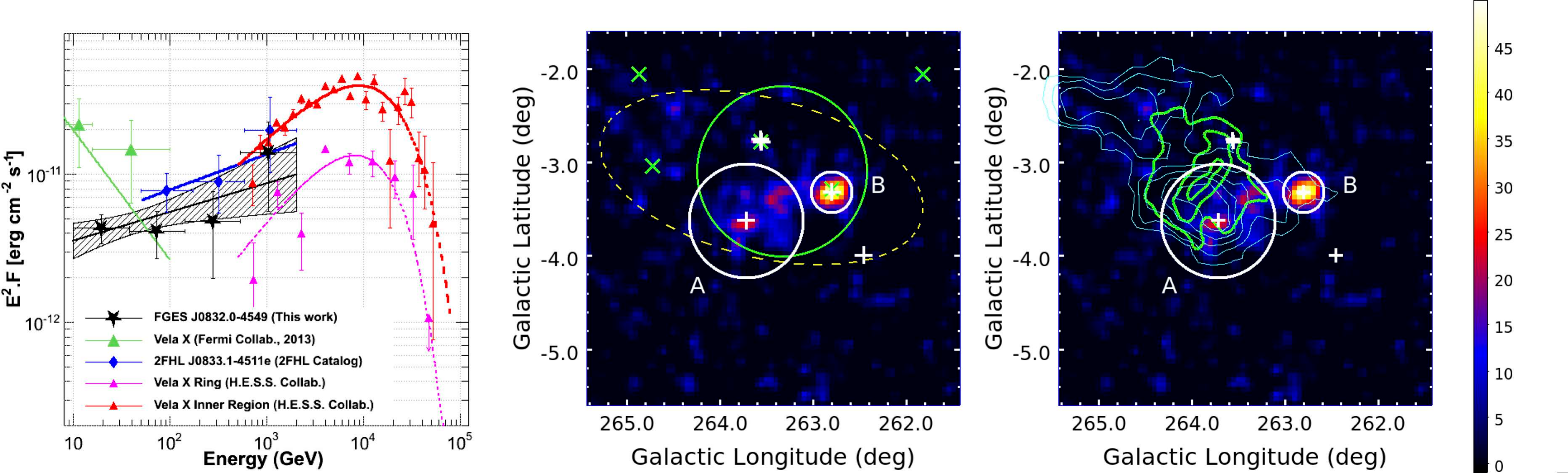}
\end{tabular}
\end{center}
\caption{
\label{fig:velax}Left: Spectral energy distributions of the pulsar wind nebula Vela-X with data points from this analysis (black stars and dashed butterfly), 2FHL catalog (blue diamonds and line), previous \emph{Fermi}-LAT publication \cite[green triangles and line,][]{2013ApJ...774..110G} and IACT data \cite[dotted red and dashed pink lines for the inner and outer emissions respectively, ][]{2012A&A...548A..38A}. Middle and right: Background-subtracted TS maps of Vela-X using the same conventions of Figure~\ref{fig:agree1} and above-quoted reference for the TeV contours shown in green. Middle: the extent of Vela-X (fit as an elliptical Gaussian) presented in the previous publication is shown with a yellow dashed ellipse. Right: the contours of the radio and IACT emission are shown in cyan \citep{2013ApJ...774..110G} and green, respectively. White circles and crosses indicate the disk extension and centroid fit in this work for Vela-X (FGES J0832.0$-$4549 labeled as A) as well as for the nearby source FGES J0830.3$-$4453 (labeled as B).
}
\end{figure*}

Previous publications on \emph{Fermi}-LAT sources used various spatial templates other than a uniform disk: Gaussian, elliptical disk, elliptical Gaussian or templates derived from multi-wavelength data. For this reason, it is not a surprise that the results presented in this work differ for such sources.\\ 
\begin{itemize}
\item {\bf HESS~J1841$-$055 (FGES~J1839.4$-$0554 and FGES~J1841.4$-$0514):}
The highly extended TeV source HESS~J1841$-$055, discovered during the H.E.S.S. Galactic Plane Survey \citep{2008A&A...477..353A}, was previously analyzed assuming the published morphology, an elliptical Gaussian with extensions of 0.41$^{\circ}$ and 0.25$^{\circ}$ \citep{2013ApJ...773...77A}. In this new work, it is detected as two separate disks whose origin and real separation remain unclear since both $\gamma$-ray components present the same spectral shape as can be seen in Figure~\ref{fig:hessj1841} (left). {One of these two components, FGES~J1841.4$-$0514, is in very good agreement with the source named Fermi J1841.1$-$0458 by \cite{2017ApJ...837...69Y}.} The H.E.S.S. source shows possibly three peaks suggesting that the emission is composed of more than one source. Several counterparts have been proposed, such as the high-mass X-ray binary system AX J1841.0$-$0536, PSR J1841$-$0524, and PSR J1838$-$0549, but none of them could solely power the whole H.E.S.S. source. More recently, the blind search detection of the new $\gamma$-ray pulsar PSR J1838$-$0537 with \emph{Fermi}-LAT \citep{2012ApJ...755L..20P} provided the only potential counterpart sufficiently energetic to power the whole H.E.S.S. source with a conversion efficiency of 0.5~\%, similar to other suggested pulsar/PWN associations. However, the spectra derived in this analysis are relatively soft in comparison to other PWNe detected at GeV energies by \emph{Fermi}-LAT, suggesting that part of the low-energy emission could have another origin. Even if the two components remain unidentified, it should be noted that the sum of their individual spectra is in very good agreement with the spectrum derived by the H.E.S.S. experiment.
\item {\bf The region of Vela-X (FGES J0830.3$-$4453 and FGES J0832.0$-$4549):}
The detection of the Vela-X PWN was reported by \emph{Fermi}-LAT in the first year of the mission and then re-investigated using four years of data, showing that it is best described by an elliptical distribution (Gaussian or disk) \citep{2013ApJ...774..110G}. This analysis also reported the detection of a significant energy break at $\sim$2 GeV in the:
\emph{Fermi}-LAT spectrum as well as a marginal spectral difference between the Northern and the Southern sides of the elliptical Gaussian. In our new analysis two sources are detected in coincidence with Vela-X (FGES J0830.3$-$4453 and FGES J0832.0$-$4549) as can be seen in Figure~\ref{fig:velax} (right). FGES J0832.0$-$4549, which is close to the cocoon as seen by H.E.S.S., has a harder spectrum consistent with the TeV points, while FGES J0830.3$-$4453 has a softer spectrum in agreement with the spectrum derived for the whole elliptical Gaussian in the former \emph{Fermi}-LAT study.
\item {\bf W44 (FGES J1856.3+0122):}
Several analyses of the middle-aged remnant W44 were performed in the GeV energy range by \emph{Fermi}-LAT and AGILE. First, using one year of \emph{Fermi}-LAT data, \cite{2010Sci...327.1103A} showed that the $\gamma$-ray source is best fit by an elliptical ring in perfect coincidence with the shell, implying that the emission is produced by particles accelerated there. Then, \cite{2012ApJ...749L..35U} announced the detection of significant emission, from the surrounding molecular cloud complex, produced by cosmic rays (CRs) that have escaped from W44. Finally, \cite{2011ApJ...742L..30G} and \cite{2013Sci...339..807A} detected the characteristic pion-decay feature in the $\gamma$-ray spectra of W44, providing the first direct evidence that cosmic-ray protons are accelerated in this shell. The use of a uniform disk in our analysis is therefore a clear simplification with respect to previous work. However, the spectrum derived is in good agreement with previous measurements showing that the bulk of the $\gamma$-ray emission is well taken into account.
\item {\bf W51C (FGES J1923.3+1408):}
W51C is another middle-aged remnant known to be interacting with a
molecular cloud. The $\gamma$-ray emission is spatially extended and best fit with an elliptical disk in agreement with the radio and X-ray extent of SNR W51C \citep{2009ApJ...706L...1A}. Recently, \cite{2016ApJ...816..100J} re-investigated the spectrum of the source down to 60 MeV and revealed a clear break at 290 MeV associated with the energy threshold of $\pi^0$ production. This result makes W51C the third unambiguously identified cosmic-ray accelerating SNR. Although the uniform disk does not perfectly reproduce the $\gamma$-ray morphology from this SNR, the spectrum is in good agreement with the previously published values. 
\item {\bf Cygnus cocoon (FGES J2026.1+4111):}
Using two years of \emph{Fermi}-LAT data, \cite{2011Sci...334.1103A} found a large excess of hard emission extending far beyond the sizes of Cyg OB2 and $\gamma$ Cygni, and following the regions bounded by photon-dominated regions as in a cocoon. The $\gamma$-ray emission peaks toward massive-star clusters and toward the southernmost molecular cloud and is well fit by a Gaussian of 2.0$^{\circ}$ width. Such a complex and highly extended region cannot be well reproduced by a simple disk.
\item {\bf The SNR S147 (FGES J0537.6+2751):}
This SNR, located toward the Galactic anticenter, is one of the most evolved SNRs in our Galaxy. No  X-ray emission  has  been  reported to  date from this region nor any TeV emission. Using 31 months of \emph{Fermi}-LAT data, \cite{2012ApJ...752..135K} reported the detection of a spatially extended $\gamma$-ray source coinciding with the SNR, with an apparent spatial correlation with prominent H$\alpha$ filaments of S147. Again, a simple disk might not be ideal to reproduce perfectly the morphology of this source, or the difference could be due to energy dependence as for W30 (see below).

\end{itemize}

\subsubsection{Energy dependence}
\begin{figure*}[ht]
\begin{center}
\begin{tabular}{ll}
\includegraphics[width=0.98\textwidth]{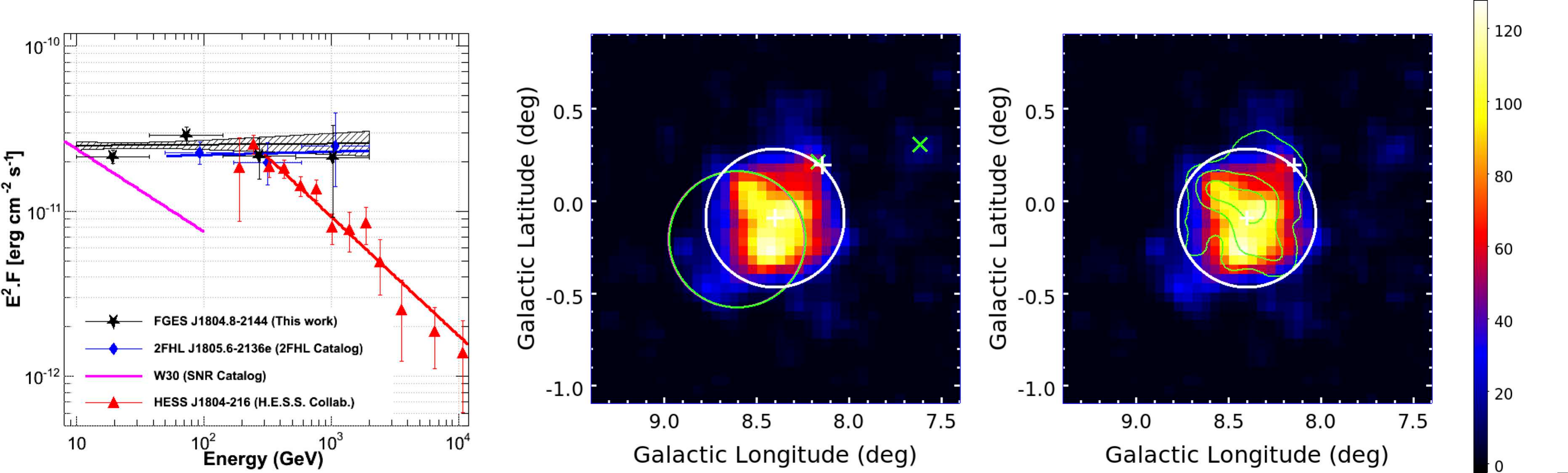}
\end{tabular}
\end{center}
\caption{
\label{fig:w30}Left: Spectral energy distributions of the star-forming region W30 by combining data from this analysis (black stars and dashed butterfly), 2FHL catalog (blue diamonds and line), SNR catalog (purple line) and IACT data \citep[red triangles and line,][]{2006ApJ...636..777A}. Middle and right: Background-subtracted TS maps of W30 using the same conventions of Figure~\ref{fig:agree1} and above-quoted reference for the IACT contours shown in green. The white circle indicates the disk extension fit in this work.
}
\end{figure*}

\begin{itemize}
\item {\bf The star-forming region W30 (FGES~J1804.8$-$2144):}
The case of the middle-aged SNR G8.7$-$0.1 located within the star-forming region W30 very well highlights the effect of energy dependence. Using 23 months of \emph{Fermi}-LAT data, \cite{2012ApJ...744...80A} detected an extended source with most of its emission in positional coincidence with the SNR G8.7$-$0.1 and a lesser part located outside the western boundary of G8.7$-$0.1. The best fit of the source morphology above 2 GeV was obtained for a disk of radius 0.37$^{\circ}$ with a reasonable correlation with the VLA radio data at 90~cm but poor correlation with the TeV data of the nearby unidentified TeV source HESS~J1804$-$216. In our new analysis, the best-fit disk has a similar radius of 0.38$^{\circ}$ but its centroid is now exactly coincident with the TeV source, providing the first evidence of an association between the GeV and TeV emissions as can be seen in Figure~\ref{fig:w30} (middle and right). It could well be that the morphological change is due to the different energy thresholds employed (2 GeV versus 10 GeV here). However, the question of the origin of the source is still unsolved. The first possibility is that the GeV and TeV emission arise from the IC scattering of the relativistic electrons in a PWN powered by the pulsar PSR J1803$-$2137. However, the relatively soft GeV spectrum (Figure ~\ref{fig:w30}, left) and large spatial extent are unusual for a PWN; the only other similar case so far is Vela-X. This would make HESS~J1804$-$216 an excellent case to investigate further since the associated X-ray PWN J1804-2140 detected by \emph{Suzaku} \citep{2007ApJ...670..643K} is not well-studied so far. The second possibility would be that GeV and TeV emissions originate from the interaction of CRs that have escaped from G8.7$-$0.1 with nearby molecular clouds. Such a scenario was proposed by \cite{2012ApJ...744...80A} to constrain the diffusion coefficient of the particles. 
\end{itemize}

\subsubsection{Improved analyses and increased statistics}
\begin{figure*}[ht]
\begin{center}
\begin{tabular}{ll}
\includegraphics[width=0.98\textwidth]{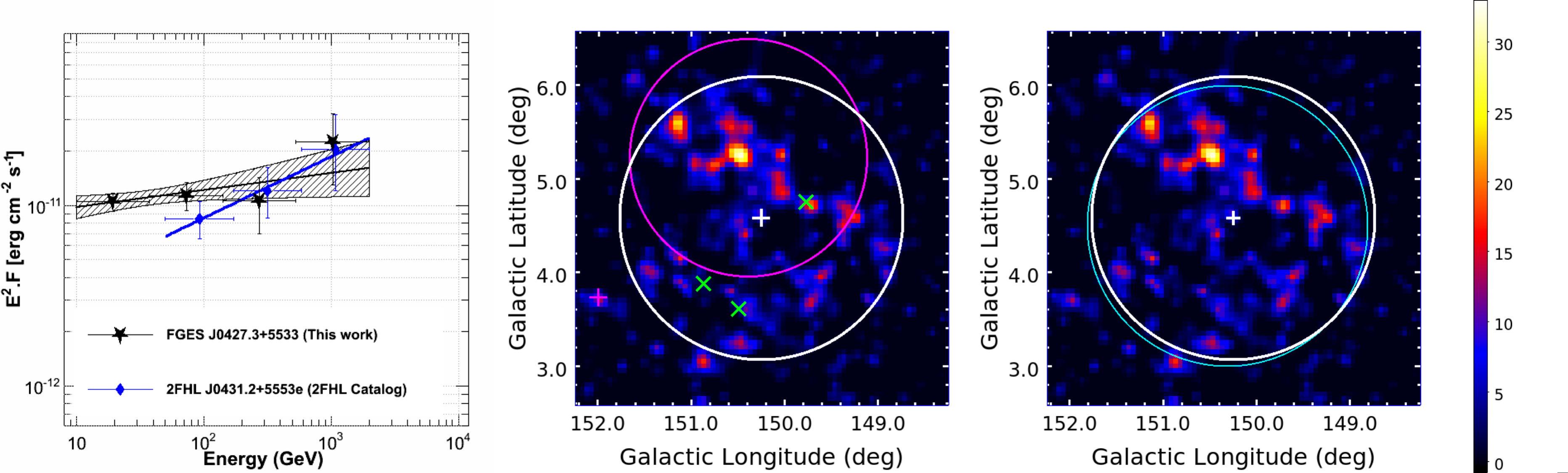}
\end{tabular}
\end{center}
\caption{
\label{fig:g150.3}Left: Spectral energy distributions of the SNR G150.3+4.5 with data points from this analysis (black stars and dashed butterfly) and 2FHL catalog (blue diamonds and line). Middle and right: Background-subtracted TS maps of SNR G150.3+4.5 using the same conventions of Figure~\ref{fig:agree1}. The white circle and central cross indicate the disk extension and centroid as fit in this work. Right : the radio extent of the SNR is shown in cyan \citep{2014A&A...567A..59G}.
}
\end{figure*}

\begin{itemize}
\item {\bf SNR G150.3+4.5 (FGES J0427.2+5533):}
The search for extended sources performed for the 2FHL catalog allowed the detection of an extended source coincident with the northern side of the faint radio SNR G150.3+4.5 \citep{2014A&A...567A..59G}. Our new analysis confirms the detection of this extended source, and thanks to the increased statistics, the \emph{Fermi}-LAT source now perfectly matches the size and location of the radio SNR, as can be seen in Figure~\ref{fig:g150.3}. The hard spectrum of this SNR derived here from 10 GeV up to 2 TeV, with $\Gamma \sim 1.9$, is more similar to that of young shell-type remnants while its large size and faintness would suggest an old age. A deeper analysis especially using \emph{Fermi}-LAT data down to 100 MeV and Cherenkov data above 2 TeV would help to constrain the characteristics of this SNR.
\end{itemize}

\subsubsection{Source confusion}
\label{subsection:confusion}
\begin{figure*}[ht]
\begin{center}
\begin{tabular}{ll}
\includegraphics[width=0.98\textwidth]{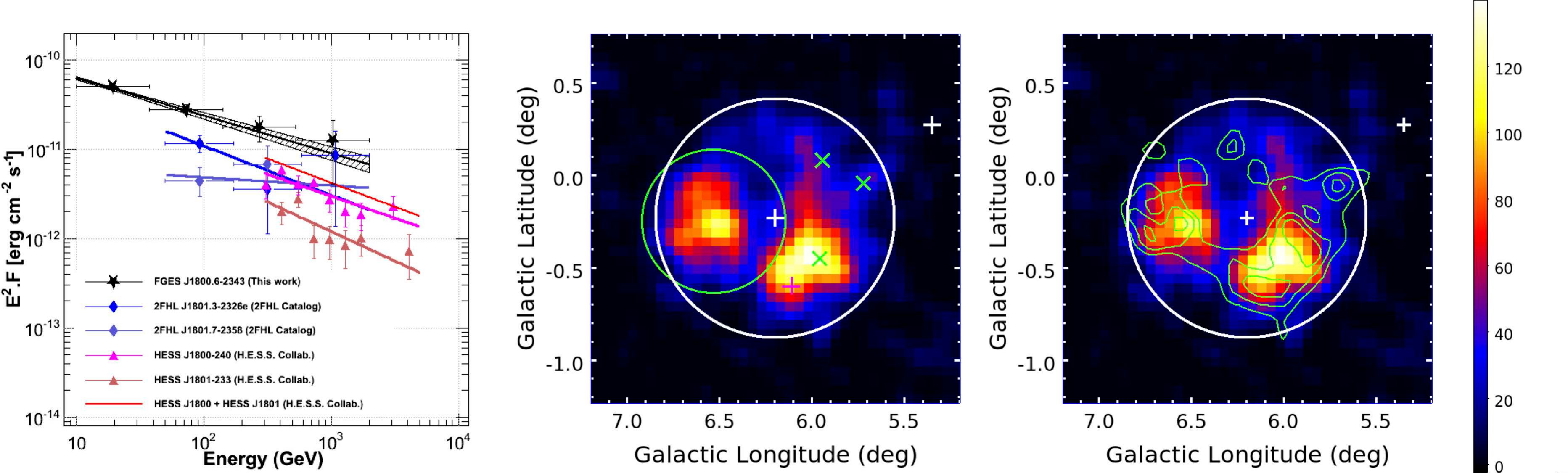}
\end{tabular}
\end{center}
\caption{
\label{fig:w28}Left: Spectral energy distributions of the SNR W28 with data points from this analysis (black stars and dashed butterfly), 2FHL catalog (blue and purple diamonds and lines for 2FHL~J1801.3$-$2326e and 2FHL~J1801.7$-$2358 respectively) and IACT data \cite[orange and pink triangles and lines for HESS~J1800$-$240 and HESS~J1801$-$233 respectively, the sum being represented in red,][]{2008A&A...481..401A}. Middle and right: Background-subtracted TS maps of SNR W28 using the same conventions of Figure~\ref{fig:agree1} and above-quoted references for the IACT contours shown in green. The white circle and central cross indicate the disk extension and centroid as fit in this work.
}
\end{figure*}

\begin{figure*}[ht]
\begin{center}
\begin{tabular}{ll}
\includegraphics[width=0.98\textwidth]{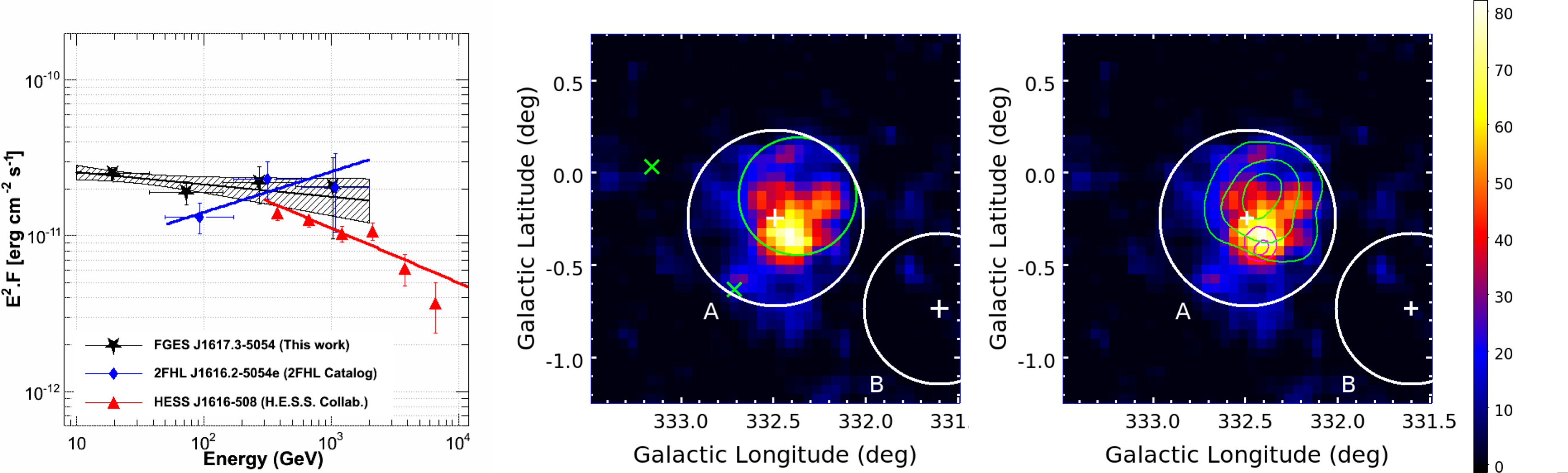}
\end{tabular}
\end{center}
\caption{
\label{fig:hessj1616}Left: Spectral energy distributions of the TeV source HESS~J1616$-$508 with data points from this analysis (black stars and dashed butterfly), 2FHL catalog (blue diamonds and line) and IACT data \cite[red line and stars, ][]{2006ApJ...636..777A}. Middle and right: Background-subtracted TS maps of HESS~J1616$-$508 using the same conventions of Figure~\ref{fig:agree1} and above-quoted references for the IACT contours shown in green. A white circle indicates the extent of the fit disk of FGES~J1617.3$-$5054 (A) and FGES~J1615.4$-$5153 (B). Right : X-rays contours (from the ROSAT All-Sky Survey) of SNR RCW~103 are overlaid in magenta.
}
\end{figure*}

\begin{figure*}[ht]
\begin{center}
\begin{tabular}{ll}
\includegraphics[width=0.98\textwidth]{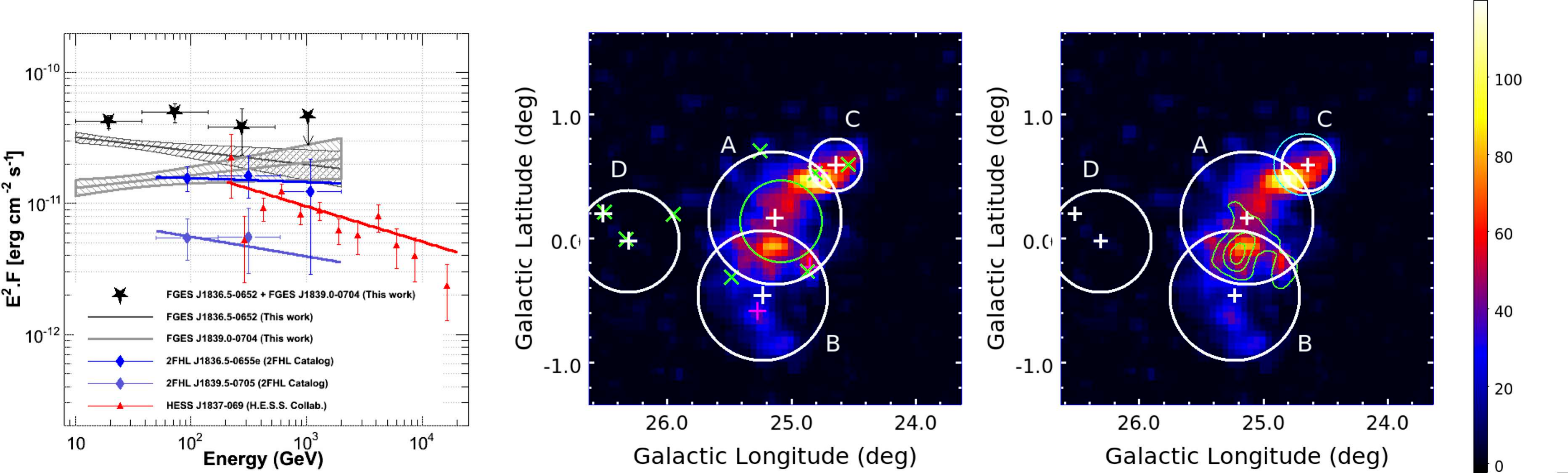}
\end{tabular}
\end{center}
\caption{
\label{fig:hessj1837}Left: Spectral energy distributions of the pulsar wind nebula HESS~J1837$-$069 with data points from this analysis (black and grey dashed butterflies for FGES J1836.5$-$0652 and FGES~J1839.0$-$0704 respectively), 2FHL catalog (blue and purple diamonds and lines for 2FHL~J1836.5$-$0655e and 2FHL~J1839.5$-$0705 respectively) and IACT data \cite[red line,][]{2006ApJ...636..777A}. Middle and right: Background-subtracted TS maps of HESS~J1837$-$069 using the same conventions of Figure~\ref{fig:agree1} and above-quoted references for the IACT contours shown in green. White circles and crosses indicate the disk extensions and centroids fit in this work for FGES~J1836.5$-$0652 (A) and FGES~J1839.0$-$0704 (B) as well as nearby extended sources FGES~J1834.1$-$0706 (C) and FGES~J1839.4$-$0554 (D).
}
\end{figure*}

\begin{figure*}[ht]
\begin{center}
\begin{tabular}{ll}
\includegraphics[width=0.98\textwidth]{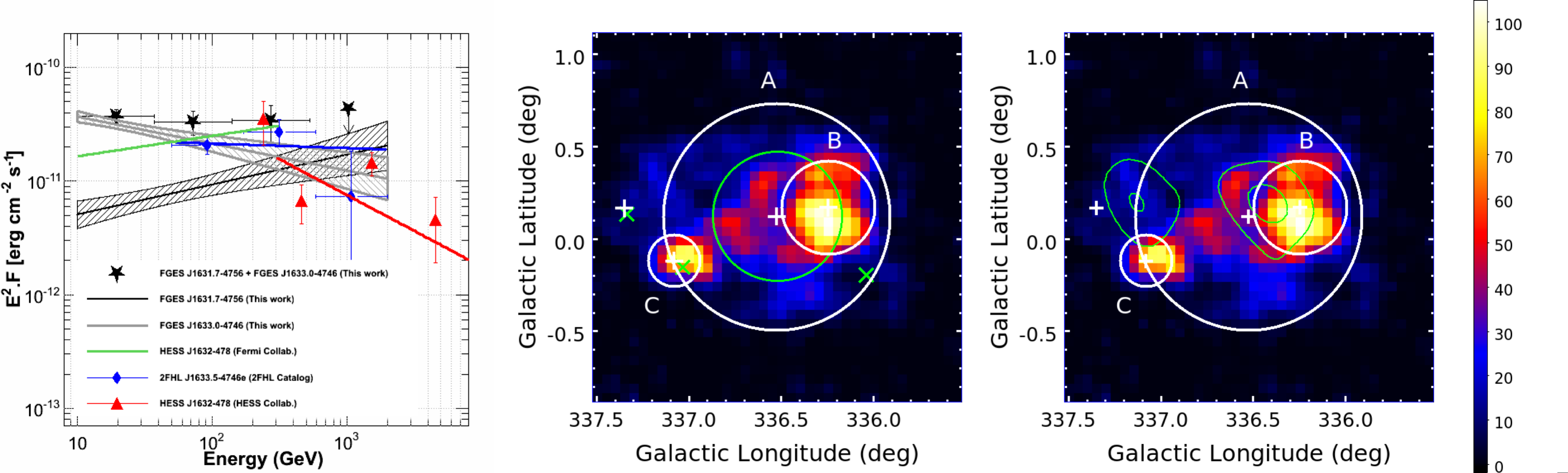}
\end{tabular}
\end{center}
\caption{
\label{fig:hessj1632}Left: Spectral energy distributions of the pulsar wind nebula HESS~J1632$-$478 with data points from this analysis (black and grey dashed butterflies for FGES~J1631.7$-$4756 and FGES~J1633.0$-$4746 respectively), 2FHL catalog (blue diamonds and line), previous \emph{Fermi}-LAT publication \cite[green line,][]{2013ApJ...773...77A} and IACT data \cite[red triangles and line, ][]{2006ApJ...636..777A}. Middle and right: Background-subtracted TS maps of HESS~J1632$-$478 and HESS~J1634$-$472 using the same conventions of Figure~\ref{fig:agree1} and above-quoted references for the IACT contours shown in green. White circles and crosses indicate the disk extensions and centroids fit in this work for FGES~J1633.0$-$4746 (A) and FGES~J1631.7$-$4756 (B) as well as the nearby extended source FGES~J1636.3$-$4731 (C).
}
\end{figure*}

Two different cases of source confusion can occur: either our extended source of interest is very close to a point source or it is near  another extended source. In such cases, the morphological fit is complex. Despite the iterative nature of our pipelines, they sometimes fail in being able to fit two nearby sources at the same time, particularly if one source is much fainter than the other. 
\begin{itemize}
\item {\bf The middle-aged SNR W28 (FGES J1800.6$-$2343):}
In the case of W28, significant $\gamma$-ray emission spatially coincident with the SNR W28 as well as the three nearby TeV sources HESS J1800$-$240A, B, and C plus another point source were detected in \cite{2014ApJ...786..145H} using 4 years of \emph{Fermi}-LAT data. The best fit of the emission coincident with W28 was obtained with a disk of 0.39$^{\circ}$ radius. In our new analysis, the disk radius of 0.64$^{\circ}$ encompasses both the SNR and the four nearby sources (see Figure~\ref{fig:w28}, middle) which explains why our disk is so large in comparison to the published value. These four sources are nearby and relatively weak which prevents a good fit of this complex region. 

\item {\bf The region of the PWN HESS J1837$-$069 (FGES J1836.5$-$0652, FGES J1839.0$-$0704):}
Within 2 degrees, this confused region contains five point sources in the 3FGL catalog in addition to the extended source associated to HESS~J1837$-$069 represented by a disk of 0.33$^{\circ}$ radius as derived by \cite{2012ApJ...756....5L}. However, the H.E.S.S. source HESS~J1837$-$069 is almost two times smaller than the \emph{Fermi}-LAT extended source and its peak emission is located on the edge of the \emph{Fermi}-LAT source. This highlights well the complexity of this region. \cite{Katsuta} re-investigated this region using 57 months of \emph{Fermi}-LAT data and detected two extended sources of $1.4^{\circ} \times 0.6^{\circ}$ and one point-like source. Doing a morphologically-resolved spectral analysis, they found that a $0.4^{\circ}$ diameter sub-region surrounding the PWN HESS J1837$-$069 has a photon index of $1.5 \pm 0.3$ while all other parts have a photon index of $2.1 \pm 0.1$ without significant spectral curvature. In this new analysis, the region is divided into three extended sources as can be seen in Figure~\ref{fig:hessj1837}: FGES J1836.5$-$0652 and FGES J1839.0$-$0704 covering HESS~J1837$-$069 and FGES J1834.1$-$0706 in the North whose size and spectrum agrees with those derived by \cite{2016ApJS..224....8A}. It is coincident with the composite SNR G24.7+0.6 and matches the radio size, supporting the association. However, the PWN HESS~J1837$-$069 can only partly explain the two extended sources FGES J1836.5$-$0652 and FGES J1839.0$-$0704 since they are much brighter and larger than the TeV signal. \cite{Katsuta} proposed a scenario in which the \emph{Fermi}-LAT emission would be produced by a star-forming region driven by a candidate young massive OB association/cluster G25.18+0.26 detected in X-ray. This would be the second case detected by the \emph{Fermi}-LAT with the Cygnus Cocoon and, indeed, they share similar spectral properties.   

\item {\bf The region of HESS~J1616$-$508 (FGES J1617.3$-$5054):}
This TeV source was detected during the H.E.S.S. Galactic plane survey \citep{2006ApJ...636..777A}. It lies in a complex region with two SNRs RCW~103 (G332.4$-$0.4) and Kes~32 (G332.4+0.1), three pulsars (PSR~J1614$-$5048, PSR~J1616$-$5109, and PSR~J1617$-$5055) and close to the SNR candidate HESS~J1614$-$518~\citep{2016arXiv161200261G} also detected in this analysis (FGES J1615.4$-$5153 in section~\ref{subsection:agree}). Only PSR~J1617$-$5055 is energetic enough to power the TeV emission of HESS J1616$-$508 and \cite{2006ApJ...636..777A} speculated that it could be a PWN powered by this young pulsar. It was detected for the first time as an extended source at GeV energies by \cite{2012ApJ...756....5L}. The disk size obtained at this time was $0.32^{\circ} \pm 0.04^{\circ} \pm 0.01^{\circ}$ which is smaller than our value of $0.48^{\circ} \pm 0.02^{\circ} \pm 0.01^{\circ}$ reported in Table~\ref{tab:morpho} as can be seen in Figure~\ref{fig:hessj1616} (middle). This discrepancy seems to be due to the contamination by the 3FGL source J1620.0$-$5101 which was removed from our sky model by our automatic pipeline but kept as a distinct source in the previous analysis. Despite this inconsistency concerning the spatial model, the agreement with the TeV spectrum is excellent.

\item {\bf The region of HESS~J1632$-$478 (FGES J1631.7$-$4756):}
The region covering the TeV PWN HESS~J1632$-$478 and unidentified source HESS J1634$-$472 is extremely complex since they are embedded in a region of the Galactic plane with bright background emission. They were both detected at GeV energies by \cite{2013ApJ...773...77A}: the source coincident with HESS J1634$-$472 was point-like whereas the source coincident with HESS~J1632$-$478 was modelled with a Gaussian distribution with a size almost twice as large as the TeV size, showing that this source might suffer from contamination. In our new analysis, an extended source (FGES J1631.7$-$4756) is detected with a relatively good match to the position and size of the TeV source HESS~J1632$-$478 and a good spectral connection with the H.E.S.S. PWN. However, no source is found coincident with HESS J1634$-$472. This might be due to the fact that we introduced a very large source (FGES J1633.0$-$4746) to take into account the bright diffuse emission in the Plane, a point source on the western edge of HESS J1634$-$472 and another slightly extended source (FGES J1636.3$-$4731) in its southern edge, as can be seen in Figure~\ref{fig:hessj1632}. This last extended source is coincident with SNR G337.0$-$0.1, which forms the CTB 33 complex together with several H II regions. Significant \emph{Fermi}-LAT emission was also found by \cite{2016ApJS..224....8A} but kept as an unidentified source due to the very large radius (0.29$^{\circ}$) obtained in comparison to the associated radio source (1.5 arcmin). Here we obtained a disk radius of 0.14$^{\circ}$, still larger than the radio shell, but coincident with it and with an OH(1720 MHz) maser spot which supports the association with the SNR as discussed by \cite{2013ApJ...774...36C}.  

\item {\bf The CTB 37 A/B complex (FGES J1714.3-3823):}
Two supernova remnants form the CTB 37 complex: the SNR CTB 37A (G348.5+0.1, associated with the TeV $\gamma$-ray source HESS J1714$-$385) and the shell-type SNR CTB 37B (G348.7+0.3, associated with HESS J1713$-$381). CTB 37A is a bright source at GeV energies and was detected by \cite{2010ApJ...717..372C} as a point source. A subsequent analysis revealed evidence for extension of $0.13^{\circ}$ at the 4.5$\sigma$ level \citep{2013AdSpR..51..247B}. Recently, \cite{2016ApJ...817...64X} announced the detection of significant \emph{Fermi}-LAT emission on CTB 37B, separated by an angular distance of less than $0.35^{\circ}$ from CTB 37A. Here, we obtained a disk radius of $0.26^{\circ}$ which encloses the whole CTB 37A/B complex. Interestingly, the value of TS$_{\rm 2pts}$ of 44 is extremely close but lower than the value of TS$_{\rm ext}$ which clearly shows that a confusion exists in this region in our analysis. In this respect, the disk radius of $0.18^{\circ} \pm 0.01^{\circ}$ obtained by \cite{2016arXiv160703778L} using \emph{Fermi}-LAT data for CTB 37A is in better agreement with the radio extension of the shell reported by \cite{1996A&AS..118..329W}.

\item {\bf The PWN HESS J1825$-$137 (FGES J1825.2$-$1359):}
This PWN is powered by the energetic radio pulsar PSR J1826$-$1334 and presents a compact core in X-rays with a hard photon index ($\Gamma = 1.6^{+0.1}_{-0.2}$) of size $30''$ embedded in a larger diffuse structure of extension $\sim5'$ extending to the south of the pulsar with a softer photon index of $\Gamma = 2.3^{+0.4}_{-0.3}$ \citep{2003ApJ...588..441G}. The TeV $\gamma$-ray emission detected by H.E.S.S. has a much larger extent ($\sim0.5^{\circ}$) but shows a similar softening of the photon index from 2.0 close to the pulsar to 2.5 at a distance of 1$^{\circ}$ \citep{2006A&A...460..365A}. The emission detected by \cite{2011ApJ...738...42G} using 20 months of \emph{Fermi}-LAT data above 1 GeV is also significantly extended with a disk radius of $0.67^{\circ} \pm 0.02^{\circ}_{\rm stat}$. Here, we obtained a larger disk radius of $1.05^{\circ} \pm 0.02^{\circ}_{\rm stat} \pm 0.25^{\circ}_{\rm syst}$ which suffers large systematics due to its location in a confused region with three bright \emph{Fermi}-LAT sources enclosed in the disk as can be seen in Figure~\ref{fig:hessj1825} (middle and right). Despite this difference of spatial model (uniform disk in this analysis with respect to a Gaussian at TeV energies), the agreement with the H.E.S.S. result is reasonable, as can be seen in Figure~\ref{fig:hessj1825} (left).

\end{itemize}

\begin{figure*}[ht]
\begin{center}
\begin{tabular}{ll}
\includegraphics[width=0.98\textwidth]{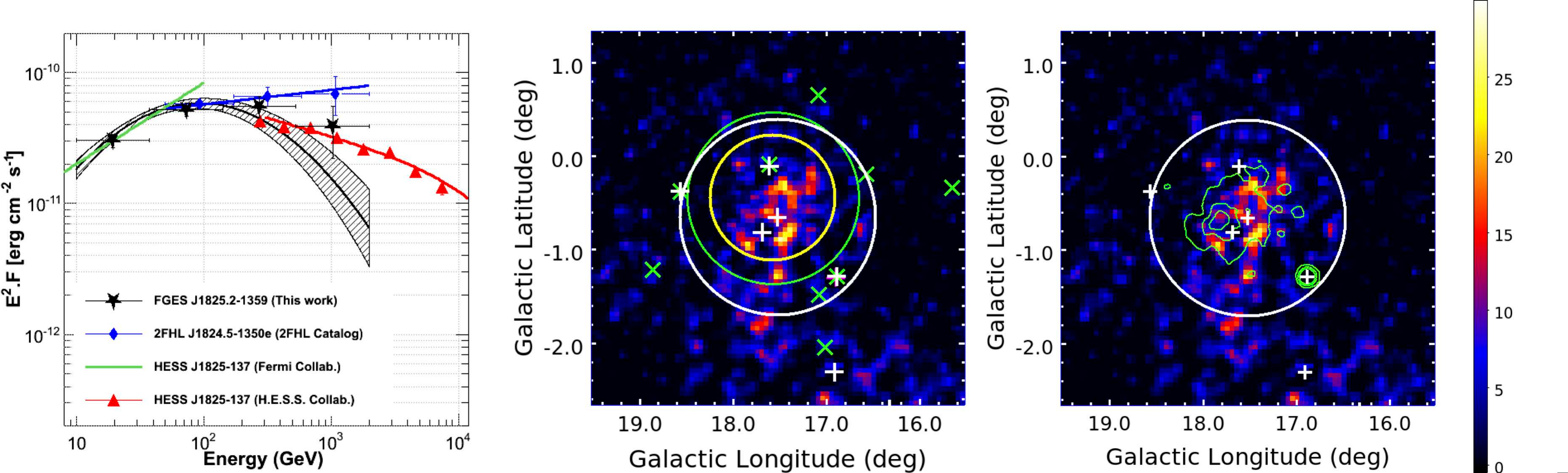}
\end{tabular}
\end{center}
\caption{
\label{fig:hessj1825}Left: Spectral energy distributions of the pulsar wind nebula HESS~J1825$-$137 by combining data from this analysis (black stars and dashed butterfly), 2FHL catalog (blue diamonds and line), a previous \emph{Fermi}-LAT publication \citep[green line,][]{2011ApJ...738...42G} and IACT data \citep[red triangles and line,][]{2006A&A...460..365A}. Middle and right: Background-subtracted TS maps of HESS~J1825$-$137 using the same conventions of Figure~\ref{fig:agree1} and above-quoted reference for the IACT contours shown in green. The white circle and cross indicate the disk extension and centroid fit in this work for FGES~J1825.2$-$1359. Middle : the extent of the disk obtained in the former $Fermi$-LAT publication is marked with a yellow circle.
}
\end{figure*}

\begin{figure*}[ht]
\begin{center}
\begin{tabular}{ll}
\includegraphics[width=0.98\textwidth]{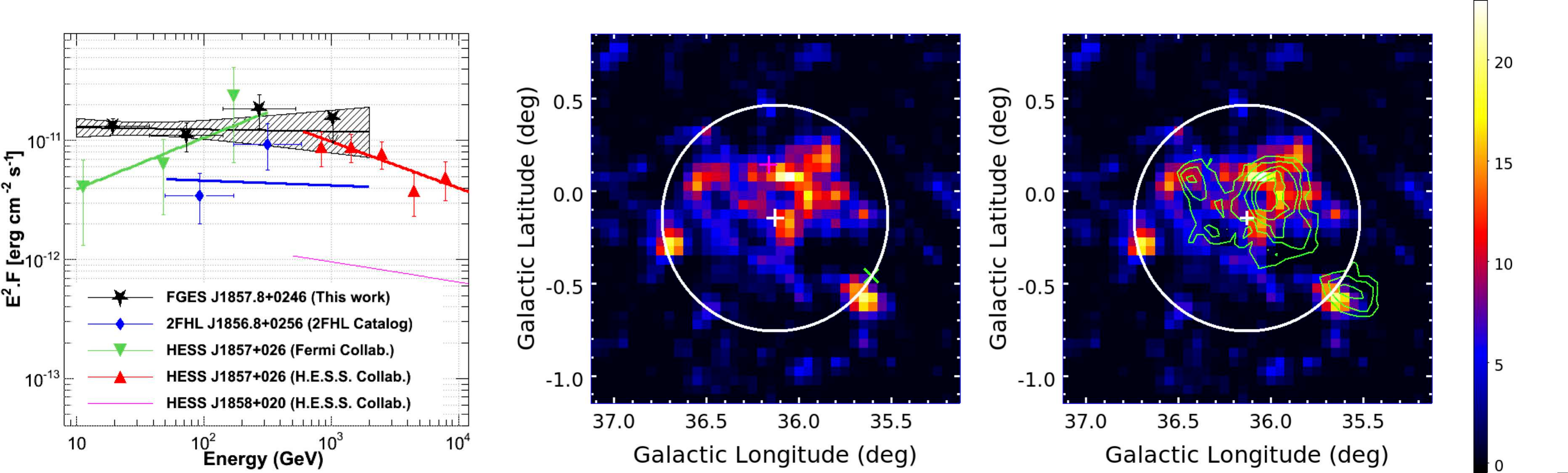}
\end{tabular}
\end{center}
\caption{
\label{fig:hessj1857}Left: Spectral energy distributions of the pulsar wind nebula HESS~J1857+026 with data points from this analysis (black line and stars, and dashed butterfly), 2FHL catalog (blue diamonds and line), previous \emph{Fermi}-LAT publication \cite[green triangles and line,][]{2012A&A...544A...3R} and IACT data \cite[red triangles and line for HESS~JJ1857+026 and pink line for HESS~J1858+020, ][]{2008A&A...477..353A}. Middle and right: Background-subtracted TS maps of HESS~J1857+026 and HESS~J1858+020 using the same conventions of Figure~\ref{fig:agree1} and above-quoted references for the IACT contours shown in green. A white circle and cross indicate the disk extension and centroid fit in this work.
}
\end{figure*}

\begin{figure*}[ht]
\begin{center}
\begin{tabular}{ll}
\includegraphics[width=0.98\textwidth]{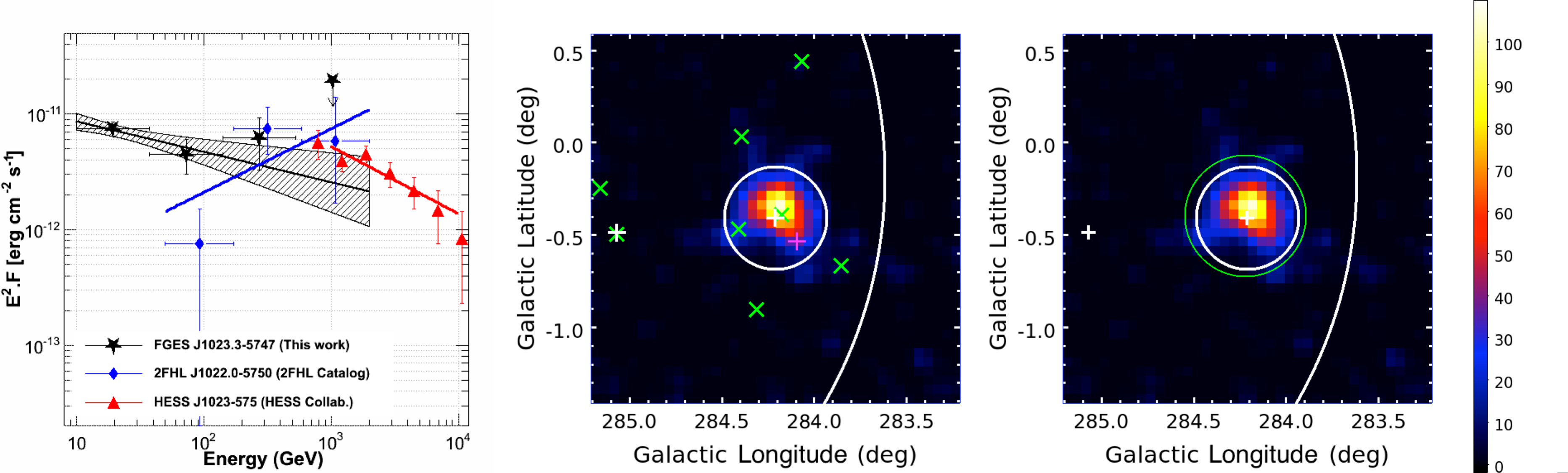}
\end{tabular}
\end{center}
\caption{
\label{fig:wd2}Left: Spectral energy distributions of HESS J1023$-$575 with data points from this analysis (black line and stars, and dashed butterfly), 2FHL catalog (blue diamonds and line) and and IACT data \cite[red triangles and line, ][]{2011A&A...525A..46H}. Middle and right: Background-subtracted TS maps of HESS J1023$-$575 using the same conventions of Figure~\ref{fig:agree1} and above-quoted references for the IACT contours shown in green. A white circle and cross indicate the disk extension and centroid fit in this work. 
}
\end{figure*}

\begin{figure*}[ht]
\begin{center}
\begin{tabular}{ll}
\includegraphics[width=0.98\textwidth]{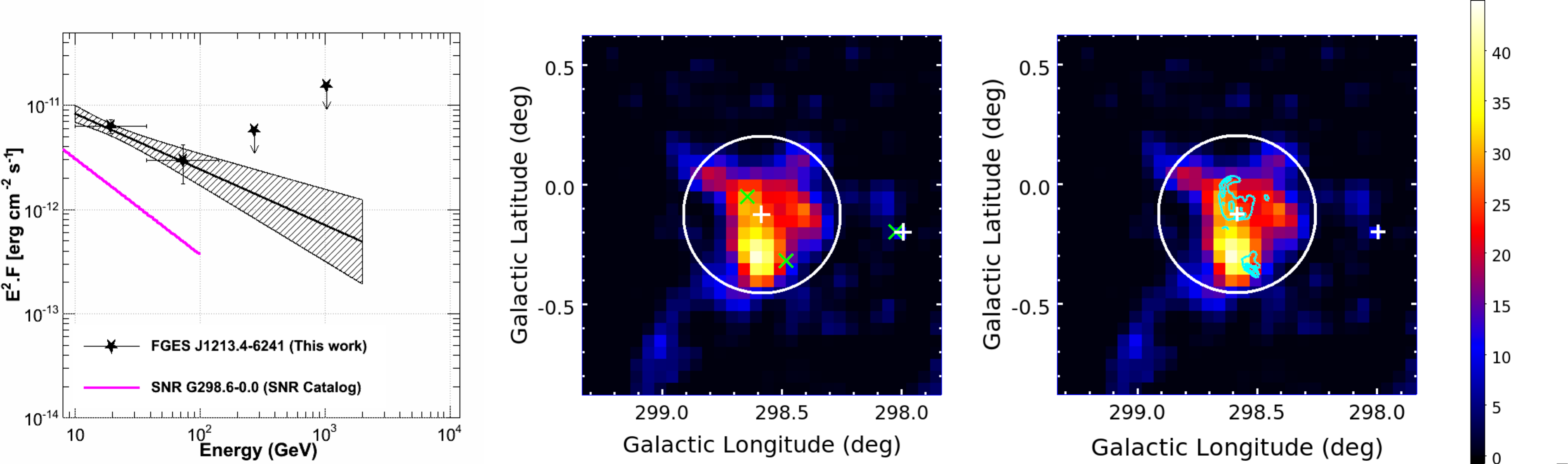}
\end{tabular}
\end{center}
\caption{
\label{fig:g298}Left: Spectral energy distributions of the SNR G298.6$-$0.0 with data points from this analysis (black line and stars, and dashed butterfly) and SNR Catalog (magenta line). Middle and right: Background-subtracted TS maps of SNR G298.6$-$0.0 using the same conventions of Figure~\ref{fig:agree1}. The radio contours of G298.6$-$0.0 and G298.5$-$0.3 (Sydney University Molonglo Sky Survey at 843 MHz, \cite{1999AJ....117.1578B}) are overlaid in cyan. A white circle and cross indicate the disk extension and centroid fit in this work.
}
\end{figure*}

\subsection{New extended sources}
\label{subsection:new}
Among the 16 new sources detected with significant extension in this analysis, eight coincide with clear counterparts and are discussed further below, except FGES J1636.3-4731 coincident with SNR G337.0$-$0.1 and already described above in Section~\ref{subsection:confusion}. 
The others seem to be confused or contaminated by the diffuse background in complex regions: FGES J1745.8$-$3028 in the Galactic Center region \citep[see][for a detailed analysis of this complex region]{2016MNRAS.457.4262H}, FGES J1036.3J1036.3$-$5834 in the region of Westerlund 2, FGES J1109.4$-$6115 in the region of MSH 11$-$62, FGES J1409.1$-$6121, FGES J1553.8$-$5325, FGES J1633.0-4746, FGES J1652.2$-$4633 and FGES J1655.6$-$4738. These confused sources can be distinguished from the others by their large disk extension and/or large systematic uncertainties.

\begin{figure*}[ht]
\begin{center}
\begin{tabular}{ll}
\includegraphics[width=0.98\textwidth]{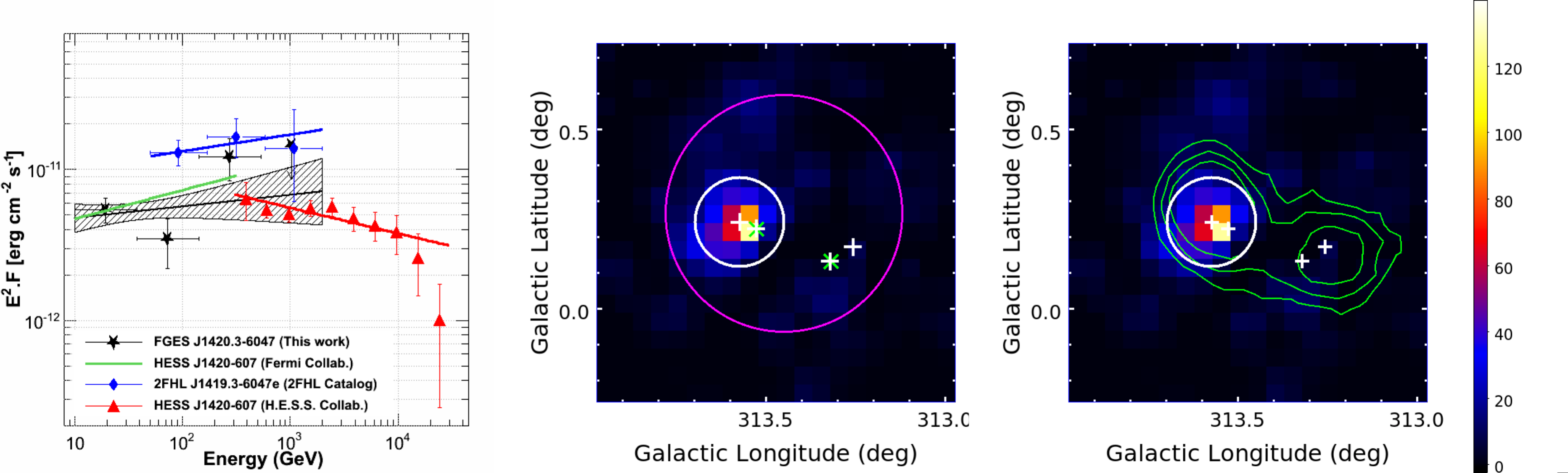}
\end{tabular}
\end{center}
\caption{
\label{fig:kooka}Left: Spectral energy distributions of the pulsar wind nebula Kookaburra with data points from this analysis (black stars and dashed butterfly), from a previous \emph{Fermi}-LAT publication \citep{2013ApJ...773...77A}, 2FHL catalog (blue diamonds and line) and  IACT data \cite[red triangles and line, ][]{2006A&A...456..245A}. Middle and right: Background-subtracted TS maps of Kookaburra using the same conventions of Figure~\ref{fig:agree1} and above-quoted references for the IACT contours shown in green. A white circle and cross indicate the disk extension and centroid fit in this work. Additional white crosses mark the positions of point sources described in the text.
}
\end{figure*}

\begin{figure*}[ht]
\begin{center}
\begin{tabular}{ll}
\includegraphics[width=0.98\textwidth]{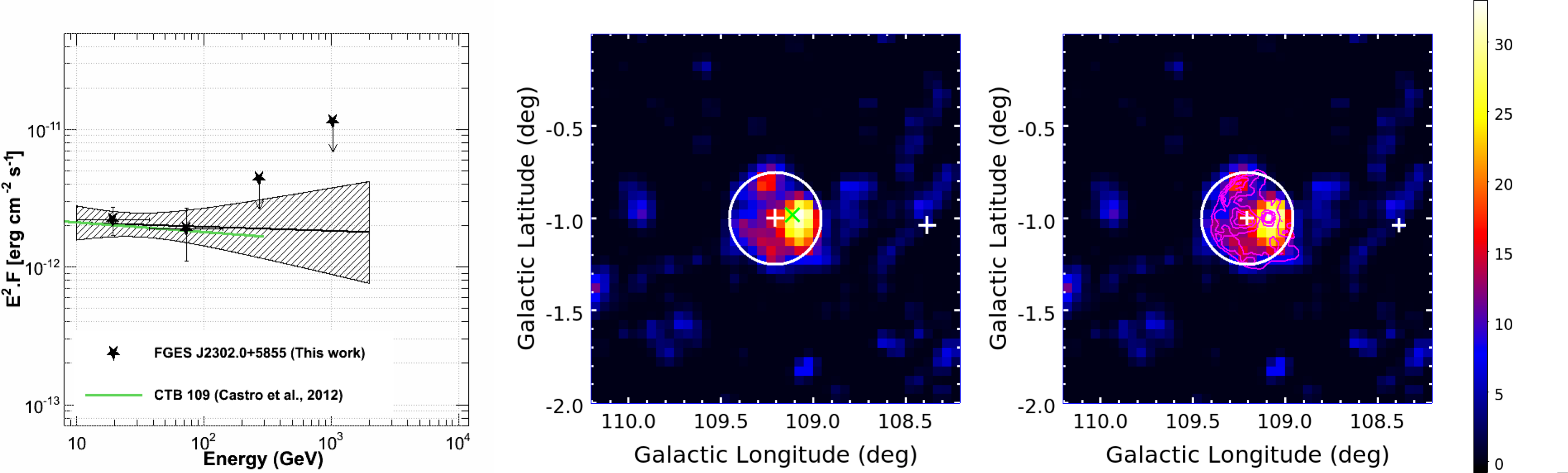}
\end{tabular}
\end{center}
\caption{
\label{fig:ctb109}Spectral energy distributions of the SNR CTB~109 constructed using data from this analysis (black stars and dashed butterfly) and from a previous publication using \emph{Fermi}-LAT data \citep[green line,][]{2012ApJ...756...88C}. Right: Background-subtracted TS maps of CTB~109 using the same conventions of Figure~\ref{fig:agree1}. A white circle and cross indicate the disk extension and centroid fit in this work. Right : X-ray contours from ROSAT PSPC (ROSAT Mission Description and Data Products Guide, available through the ROSAT Guest Observer Facility, NASA GSFC) are overlaid in magenta. 
}
\end{figure*}

\begin{figure*}[ht]
\begin{center}
\begin{tabular}{ll}
\includegraphics[width=0.98\textwidth]{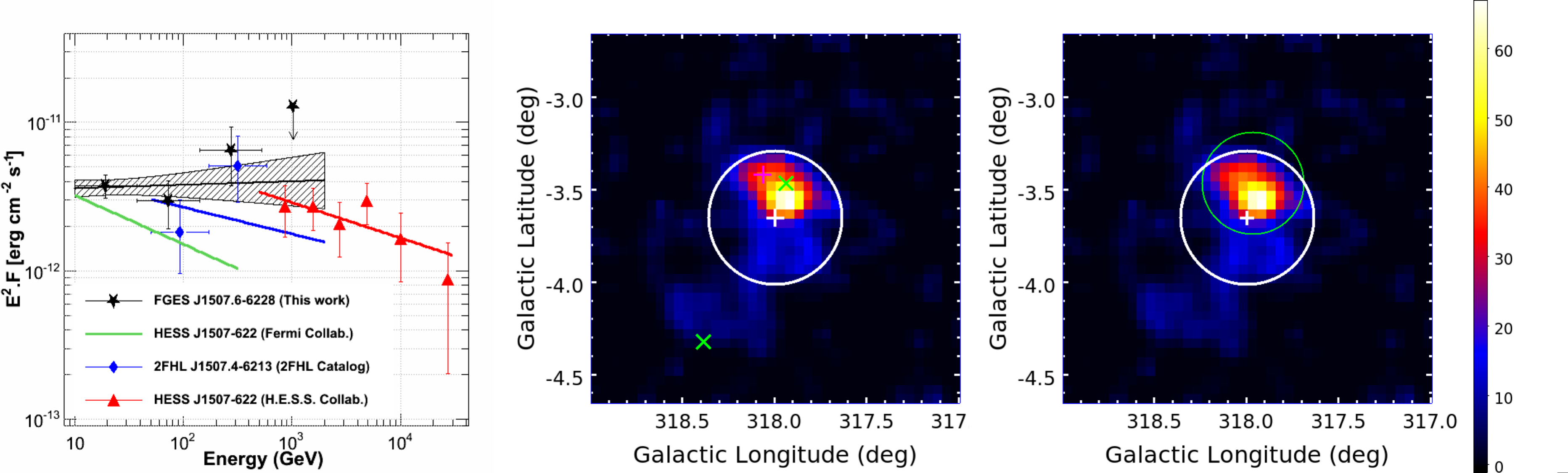}
\end{tabular}
\end{center}
\caption{
\label{fig:hessj1507}Left: Spectral energy distributions of the TeV source HESS~J1507$-$622 with data points from this analysis (black stars and dashed butterfly), from the 2FHL catalog (blue diamonds and line), a previous \emph{Fermi}-LAT publication \cite[green line,][]{2013ApJ...773...77A} and IACT data \cite[red triangles and line,][]{2011A&A...525A..45H}. Middle and right: Background-subtracted TS maps of HESS~J1507$-$622 using the same conventions of Figure~\ref{fig:agree1} and above-quoted references for the TeV extent shown in green. A white circle and cross indicate the disk extension and centroid fit in this work.
}
\end{figure*}

\begin{figure*}[ht]
\begin{center}
\begin{tabular}{ll}
\includegraphics[width=0.98\textwidth]{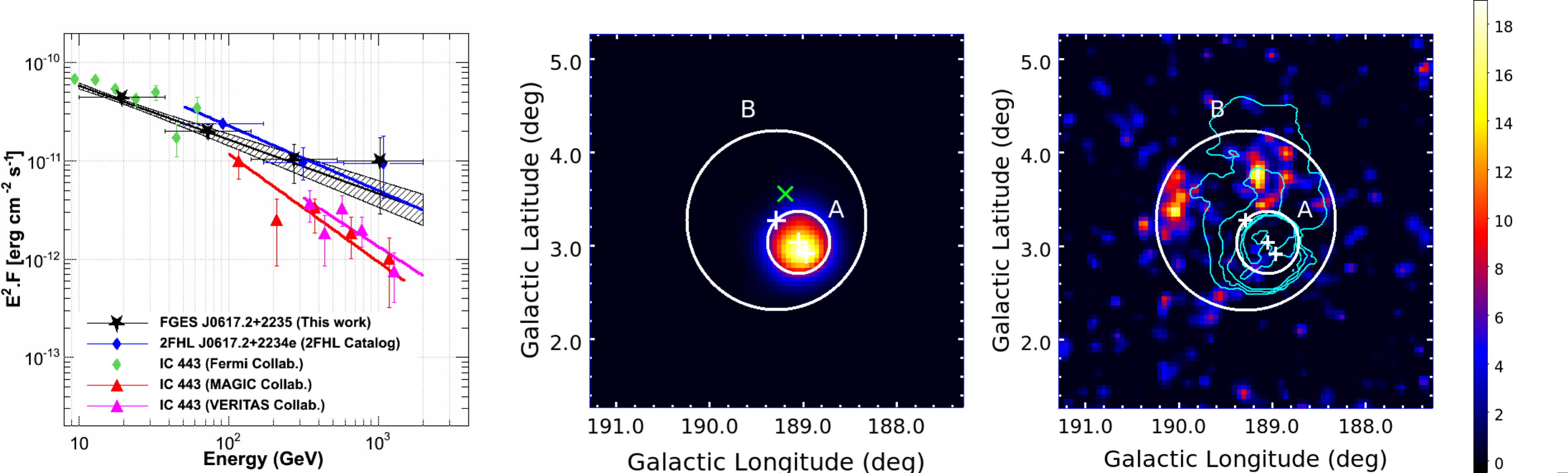}
\end{tabular}
\end{center}
\caption{
\label{fig:ic443}Left: Spectral energy distributions of the SNR IC~443 (FGES~J0617.2+2235) with data points from this analysis (black stars and dashed butterfly), from the 2FHL catalog (blue diamonds and line), a previous \emph{Fermi}-LAT publication \cite[green diamonds,][]{2013Sci...339..807A} and IACT data (red and pink triangles are taken from \cite{2007ApJ...664L..87A} and \cite{2009ApJ...698L.133A} respectively). Middle and right: Background-subtracted TS maps of IC~443 using the same conventions of Figure~\ref{fig:agree1}. The 3FGL and 2FHL source for the SNR IC 443 are exactly coincident with our FES source J0617.2+2235. White circles and crosses indicate the disk extensions and centroids fit in this work for FGES~J0617.2+2235 (A) and FGES~J0619.6+2229 (B). Right : the bright emission from FGES~J0617.2+2235 is included in the model to highlight the emission coming from the largest source FGES~J0619.6+2229. Cyan contours represent the radio emission at 1420 MHz \citep{2004AJ....127.2277L}.
}
\end{figure*}

\begin{itemize}
\item {\bf The PWN HESS J1857+026 (FGES J1857.8+0246):}
HESS~J1857+026 is a TeV $\gamma$-ray source detected by H.E.S.S. during the Galactic Plane Survey \citep{2008A&A...477..353A}. The extended ($\sim0.11^{\circ}$) TeV source was identified as a PWN candidate after the discovery of PSR J1856+0245 in the Arecibo PALFA survey by \cite{2008ApJ...682L..41H}. Recently, MAGIC reported a measured Gaussian extension in the 0.2 -- 1 TeV energy range significantly larger ($0.20^{\circ}$, equivalent to a disk width of $0.37^{\circ}$) than the extension reported by H.E.S.S. \citep{2011ICRC....7..173K, 2014A&A...571A..96M}. They demonstrated that, above 1 TeV, the emission is due to two spatially distinct statistically significant components: the extended PWN powered by PSR J1856+0245 and an unidentified point source. HESS J1857+026 was detected at GeV energies but as a single point source \citep{2010arXiv1011.0210N, 2012A&A...544A...3R}. Here, we detect an extended source coincident with 
HESS J1857+026 but with a disk extension of $0.61^{\circ} \pm 0.03^{\circ} \pm 0.06^{\circ}$ much larger than the MAGIC one. However, looking at the TS maps in Figure~\ref{fig:hessj1857}, one clearly sees two hotspots in the south-east and south-west (coincident with HESS J1858+020) part of the source which could explain the larger size reported in this analysis since they may not be related to the PWN. This does not prevent consistency between the spectra derived at GeV and TeV energies as can be seen in Figure~\ref{fig:hessj1857} (left), partly thanks to the low TeV flux of HESS J1858+020. But, it could explain the flat GeV spectrum derived in this analysis in comparison to the previously published one which used the H.E.S.S. morphology as a template \citep{2012A&A...544A...3R}.
\item {\bf The region of Westerlund 2 (FGES J1023.3-5747 and FGES J1036.3-5834):}
In 2007, H.E.S.S. reported the detection of $\gamma$-rays from an extended source of Gaussian width $0.18^{\circ}$, HESS J1023$-$575, in the direction of the young stellar cluster Westerlund 2 \citep{2007A&A...467.1075A, 2011A&A...525A..46H}. HESS J1023$-$575 was detected at GeV energies by the \emph{Fermi}-LAT but no extension was reported \citep{2011ApJ...726...35A}. In parallel, the \emph{Fermi}-LAT collaboration announced the detection of the pulsar PSR J1022$-$5746, suggesting that it could be a potential counterpart of the TeV source \citep{2010ApJ...725..571S}. Here, we report significant extension from HESS J1023$-$575 with a disk radius of $0.28^{\circ} \pm 0.02^{\circ} \pm 0.06^{\circ}$, in excellent agreement with the TeV one. The origin of the signal from HESS J1023$-$575 remains unsolved despite the new morphological and spectral results reported in this paper and illustrated in Figure~\ref{fig:wd2}. The $\gamma$-ray emission could originate from a PWN associated with PSR J1022$-$5746 or mechanisms related to acceleration of CRs in the open cluster Westerlund 2. However, the region is confused at GeV energies (with an extremely large source FGES J1036.3$-$5834 covering $2.5^{\circ}$ surrounding our source of interest) and the spectrum derived here might suffer from contamination especially around 10 GeV. A dedicated analysis is clearly needed to constrain the origin of the $\gamma$-ray signal. 
\item {\bf The SNR G298.6$-$0.0 and G298.5$-$0.3 (FGES J1213.3$-$6240):}
The SNRs G298.6$-$0.0 and G298.5$-$0.3 are both detected at 408 MHz and 843 MHz with flat radio photon index of $\sim$1.3 \citep{1987A&A...183..118K}. The possible interaction with a high-density medium from the direction of these two
SNRs was reported by \cite{2006AJ....131.1479R}, making these sources excellent targets for GeV observations. Indeed, the GeV detection of a point source coincident with the shell of G298.6$-$0.0 and G298.5$-$0.3 was reported by \cite{2015ApJS..218...23A, 2016ApJS..224....8A}. Recently, X-ray observations by \emph{Suzaku} revealed a center-filled structure inside the radio shell~\citep{2016PASJ...68S...5B}. This classifies G298.6$-$0.0 as a new mixed-morphology SNR such as IC 443 \citep{2008A&A...485..777T}. In this work, we report a significant extension at a position coincident with SNR G298.6$-$0.0 and with a size including G298.5$-$0.3, as can be seen in Figure~\ref{fig:g298}. The soft $\gamma$-ray spectrum is consistent with the fact that these sources have a spectral break around a few GeV \citep{2015ApJS..218...23A, 2016ApJS..224....8A} which is similar to most SNRs interacting with molecular clouds. The higher flux reported in this analysis can be explained by the fact that we are adding the flux of both SNRs. This makes the extension measure reported here questionable.
\item {\bf The Kookaburra complex (FGES J1420.3-6047):}
The complex of compact and extended radio/X-ray sources, called
Kookaburra \citep{1999ApJ...515..712R}, spans over one square degree along the Galactic Plane. It contains two young and energetic pulsars PSR J1420$-$6048 and PSR J1418$-$6058 powering the PWNe called ``K3" and the ``Rabbit", respectively. The H.E.S.S. Galactic Plane survey revealed two sources in this region: HESS J1420$-$607 centered north of PSR J1420$-$6048 (near K3) and HESS J1418$-$609 coincident with the Rabbit nebula \citep{2006A&A...456..245A}. In a previous analysis of the region above 10 GeV using \emph{Fermi}-LAT data~\citep{2013ApJ...773...77A}, HESS J1420$-$607 and HESS J1418$-$609 were detected as two point sources with different spectral shapes: a hard spectrum for the first one (suggesting a PWN origin) and a soft spectrum with an energy cut-off at a few GeV for the second, suggestive of pulsar emission and thus likely due to contamination from PSR~J1418$-$6058. It was then detected as a very extended source~\footnote{A typo was recently discovered in the disk extension value reported in Table 5 of \cite{2016ApJS..222....5A} and in its associated fits file. An erratum is being prepared quoting a value of 0.33$^\circ$ for this source instead of 0.36$^\circ$.} of $0.33^{\circ}$ covering both PWNe in \cite{2016ApJS..222....5A}. In our new analysis, HESS J1420$-$607 is detected as an extended source with a disk radius of 0.12$^{\circ}$ in good agreement with the TeV size, while HESS J1418$-$609 remains point-like. In addition to these two PWNe, the model of the region contains two sources coincident with their associated pulsars PSR J1420$-$6048 and PSR J1418$-$6058, as can be seen in Figure~\ref{fig:kooka}.
\item {\bf CTB 109 (FGES J2302.0+5855):}
CTB 109 (G109.1$-$1.0) is a Galactic SNR with a hemispherical shell morphology in X-rays and in the radio band. Using 37 months of \emph{Fermi}-LAT data, \cite{2012ApJ...756...88C} detected a $\gamma$-ray source coincident with the position of the remnant with no sign of significant extension. Thanks to the excellent angular resolution offered by the new Pass 8 data, the extension of the \emph{Fermi}-LAT source is now significant and in perfect agreement with the size of the remnant, ruling out an association with the giant molecular cloud located to the west of the SNR because it is too far from the centroid of the $\gamma$-ray emission. The spectrum derived in this new analysis, presented in Figure~\ref{fig:ctb109} (left) is consistent with the former one and can be reasonably fit in both leptonic and hadronic models. It should be noted that the spectrum and morphology derived here are in perfect agreement with those published recently by \cite{2016arXiv160703778L}. 
\item {\bf HESS~J1507$-$622 (FGES J1507.6-6228):}
Most $\gamma$-ray sources in the inner Galaxy H.E.S.S. survey tend to cluster within $1^{\circ}$ in latitude around the Galactic plane. HESS~J1507$-$622 instead is unique, since it is located at a latitude of $\sim$3.5$^{\circ}$ and does not have any obvious counterpart in other multi-wavelength data. Up to now, the nature of this slightly extended source (with a Gaussian width of 0.15$^{\circ} \pm 0.02^{\circ}$) is still unidentified. HESS J1507$-$622 was detected in the \emph{Fermi}-LAT energy range as a point source \citep{2012A&A...545A..94D} with a rather flat spectrum from the GeV to the TeV regime. Our new analysis confirms the former spectrum and shows for the first time a significant extension in the GeV regime, in agreement with the TeV size (see Figure~\ref{fig:hessj1507}). These results challenge an extragalactic origin due to the large energetics needed to power the source and the very extended nature of the emission in such a scenario. For a Galactic origin, the compactness of the source suggests a distance to the object of several kpc and its location far off plane may indicate a parent stellar population as old as 1 Gyr. This does not rule out a PWN origin for the source but implies a very low magnetic field of $\sim1 \mu$G to be able to explain the absence of an X-ray counterpart.
\item {\bf The region of IC 443 (FGES J0617.2+2235 and FGES J0619.6+2229):}
The middle-aged SNR IC 443 has been extensively studied at all wavelengths and established as a strong $\gamma$-ray source extended in the TeV band \citep{2007ApJ...664L..87A} and in the GeV domain \citep{2010ApJ...710L.151T, 2010ApJ...712..459A}. The $\gamma$-ray data was interpreted by \cite{2010MNRAS.408.1257T} in the framework of cosmic-ray interactions with a giant molecular cloud lying in front of the remnant. Then, using \emph{Fermi}-LAT data down to 60 MeV, \cite{2013Sci...339..807A} detected a spectral break at low energy, characteristic of pion-decay emission, proving that protons are indeed being accelerated in this remnant. More recently, \cite{2015arXiv151201911H} showed that the TeV emission as seen by VERITAS is strongly correlated with the GeV morphology of the \emph{Fermi}-LAT and extends over the entire surface of the remnant. Here our analysis finds a best disk radius of $0.34^{\circ}$ directly matching the bright northeast half-shell of $\sim$40 arcmin diameter with a good spectral agreement with previous publications. Even more interesting is the diffuse source FGES J0619.6+2229 which overlaps with IC 443 (see Figure~\ref{fig:ic443}) and extends to the North towards the bright arc and H~II region S249 seen at 1420 MHz \citep{2004AJ....127.2277L}. This source of almost $1^{\circ}$ radius presents a harder spectrum than IC~443 and may be produced by cosmic rays accelerated by the shell of IC 443 and diffusing in the surrounding medium. It could also have a different origin with a connection to the SNR G189.6+3.3 which presents non-thermal emission in radio and X-rays \citep{1994A&A...284..573A, 2004AJ....127.2277L}.
\end{itemize}

\begin{figure*}[ht]
\begin{center}
\begin{tabular}{ll}
\includegraphics[width=0.98\textwidth]{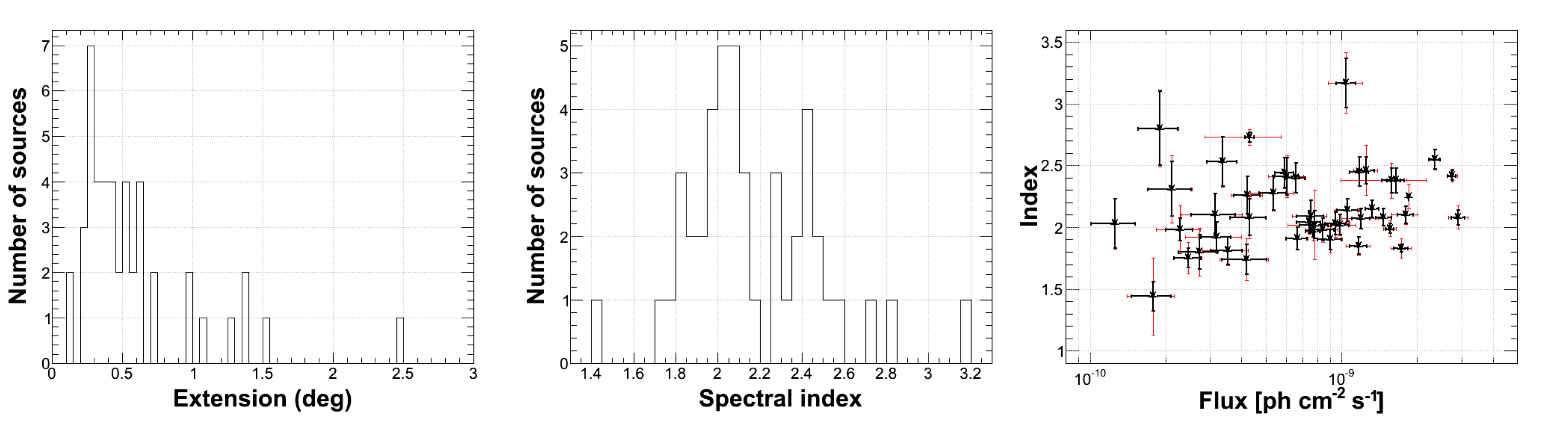}
\end{tabular}
\end{center}
\caption{
\label{fig:fin}
Distribution of the disk extension ({\it left panel}), spectral index ({\it middle panel}) and spectral index versus flux ({\it right panel}) of the 46 extended sources detected in this analysis. FGES J1825.2$-$1359 is not included in the middle and right plots since it is the only source modeled using a LogParabola. In the right panel, statistical errors are indicated in black, while the quadrature sum of systematic and statistical errors are reported in red.
}
\end{figure*}

\section{Summary}
\label{sec:summary}
Using the new Pass~8 \emph{Fermi}-LAT data above 10 GeV, we have detected 46 significantly extended sources in the Galactic plane and provided their morphological and spectral characteristics. Only four Galactic sources already detected as significantly extended in previous works were not detected in this work; none of them show any significant emission above 10 GeV in the 3FHL catalog either. These extended sources have $>$10\,GeV fluxes ranging from $\sim$$1.2 \times 10^{-10}$ to $\sim$$29\times 10^{-10}$~cm$^{-2}$~s$^{-1}$ with a median flux of 9.6$\times 10^{-10}$~cm$^{-2}$~s$^{-1}$. On average, they display hard spectra with a median spectral index of 2.1, 70\% of all sources having a spectrum harder than 2.2 and even harder than 2.0 for 40\% (see Figure~\ref{fig:fin}, right) implying a high-energy SED peak in the TeV band. The measured disk extensions show a large dispersion with values ranging from 0.12$^{\circ}$ to 2.5$^{\circ}$, with a median value of 0.5$^{\circ}$ as can be seen from Figure~\ref{fig:fin} (left).\\ 
Among the 46 extended sources, 16 are new, 13 agree with previous publications and 17 have a different morphology. This perfectly highlights that, thanks to the improved performance offered by the Pass~8 data and the increased exposure, we detect more sources and better characterize the morphology of already known sources. This is particularly evident for the case of the large SNR, G150.3+4.5 whose $\gamma$-ray morphology now perfectly matches the radio size and location. This source is not an isolated case since SNRs are the dominant class of extended sources detected in this search. All extended sources identified with PWNe are also detected at TeV energies. This may be due to a higher energy SED peak for these sources in comparison to SNRs. It is important to note that 7 SNRs and 1 star-forming region are undetected at TeV energies highlighting the excellent sensitivity of the \emph{Fermi}-LAT in the $>$ 10 GeV band thanks to its uniform exposure over the whole Galactic plane and its low background in comparison to Cherenkov telescopes. The current and future observations of the \emph{Fermi}-LAT are thus crucial to probe the $> \, 10$ GeV sky and especially the Galactic plane, providing excellent targets for current and future Cherenkov telescopes such as CTA. \\

\newpage
\begin{deluxetable*}{ccccccc}
\tablecaption{Best-fitting morphological parameters of the extended sources detected above 10 GeV\label{tab:morpho}}
\tablenum{1}
\tablehead{
\colhead{FGES Name} &
\colhead{TS} &
\colhead{TS$_{\rm ext}$} &
\colhead{TS$_{\rm 2pts}$} &
\colhead{Right Ascension} &
\colhead{Declination} &
\colhead{Extension $\sigma$} \\
\colhead{} & \colhead{} & \colhead{} & \colhead{} & \colhead{(\degrees)} & \colhead{(\degrees)} & \colhead{(\degrees)}
}
\startdata
J0427.2+5533		& 192 	& 160 	& 21 		& 66.822 $\pm$ 0.043		& 55.552 $\pm$ 0.053		& 1.52 $\pm$ 0.03 $\pm$ 0.08		\\
J0537.6+2751		& 35 		& 17 		& 9 		& 84.407 $\pm$ 0.057		& 27.859 $\pm$ 0.056		& 1.39 $\pm$ 0.02 $\pm$ 0.09		\\
J0617.2+2235		& 831 	& 572 	& 177 	& 94.309 $\pm$ 0.008		& 22.584 $\pm$ 0.008		& 0.34 $\pm$ 0.01 $\pm$ 0.01		\\
J0619.6+2229		& 68 		& 49 		& 21 		& 94.653 $\pm$ 0.040		& 22.484 $\pm$ 0.028		& 0.96 $\pm$ 0.03 $\pm$ 0.05	\\
J0822.1$-$4253		& 374 	& 198 	& 193 	& 125.545 $\pm$ 0.016		& $-$42.888 $\pm$ 0.019		& 0.44 $\pm$ 0.01 $\pm$ 0.01		\\
J0830.3$-$4453		& 80 		& 23 		& 7 		& 127.588 $\pm$ 0.041		& $-$44.885 $\pm$ 0.025		& 0.22 $\pm$ 0.03 $\pm$ 0.02	\\
J0832.0$-$4549		& 69 		& 41 		& 0 		& 128.008 $\pm$ 0.044		& $-$45.813 $\pm$ 0.046		& 0.61 $\pm$ 0.04 $\pm$ 0.05 (6/8)	\\
J0851.9$-$4620		& 808 	& 728 	& 394 	& 132.987 $\pm$ 0.021		& $-$46.342 $\pm$ 0.016		& 0.98 $\pm$ 0.01 $\pm$ 0.01	\\
J1023.3$-$5747		& 144 	& 41 		& 20 		&155.828 $\pm$ 0.021		& $-$57.794 $\pm$ 0.024		& 0.28 $\pm$ 0.02 $\pm$ 0.06	\\
J1036.3$-$5834$\dagger$ & 281 	& 265 	& 18 		& 159.094 $\pm$ 0.049		& $-$58.563 $\pm$ 0.042		& 2.47 $\pm$ 0.06 $\pm$ 0.06	\\
J1109.4$-$6115$\dagger$		& 141 	& 134 	& 24 		& 167.362 $\pm$ 0.046		& $-$61.259 $\pm$ 0.042		& 1.27 $\pm$ 0.03 $\pm$ 0.08	\\
J1213.3$-$6240		& 105 	& 66 		& 34 		& 183.346 $\pm$ 0.014		& $-$62.688 $\pm$ 0.032		& 0.33 $\pm$ 0.03 $\pm$ 0.05		\\
J1303.5$-$6313		& 93 		& 30 		& 15 		& 195.876 $\pm$ 0.021		& $-$63.224 $\pm$ 0.023		& 0.33 $\pm$ 0.02 $\pm$ 0.01		\\
J1355.1$-$6420		& 84 		& 41 		& 11 		& 208.802 $\pm$ 0.026		& $-$64.345 $\pm$ 0.023		& 0.41 $\pm$ 0.02 $\pm$ 0.01		\\
J1409.1$-$6121$\dagger$		& 237 	& 152 	& 23 		& 212.285 $\pm$ 0.020		& $-$61.355 $\pm$ 0.022		& 0.73 $\pm$ 0.02 $\pm$ 0.06		\\
J1420.3$-$6047		& 77 		& 32 		& 26 		& 215.082 $\pm$ 0.013		& $-$60.782 $\pm$ 0.011		& 0.12 $\pm$ 0.01 $\pm$ 0.01		\\
J1443.2$-$6227		&  122 	& 85 		& 18 		& 220.797 $\pm$ 0.025		& $-$62.460 $\pm$ 0.024		& 0.37 $\pm$ 0.02 $\pm$ 0.01	\\
J1507.6$-$6228		& 104 	& 34 		& 17 		& 226.984 $\pm$ 0.031		& $-$62.467 $\pm$ 0.024		& 0.36 $\pm$ 0.02 $\pm$ 0.03	\\
J1514.3$-$5910		& 517 	& 233 	& 135 	& 228.572 $\pm$ 0.014		& $-$59.163 $\pm$ 0.012		& 0.24 $\pm$ 0.01 $\pm$ 0.01	\\
J1552.9$-$5610		& 435 	& 142 	& 39 		& 238.219 $\pm$ 0.014		& $-$56.166 $\pm$ 0.015		& 0.25 $\pm$ 0.01 $\pm$ 0.01	\\
J1553.8$-$5325$\dagger$		& 192 	& 154 	& 17 		& 238.456 $\pm$ 0.024		& $-$53.424 $\pm$ 0.026		& 0.52 $\pm$ 0.02 $\pm$ 0.09	\\
J1615.4$-$5153		& 302 	& 242 	& 79 		& 243.849 $\pm$ 0.021		& $-$51.881 $\pm$ 0.024		& 0.41 $\pm$ 0.02 $\pm$ 0.06	\\
J1617.3$-$5054		& 294 	& 214 	& 37 		& 244.328 $\pm$ 0.021		& $-$50.909 $\pm$ 0.019		& 0.48 $\pm$ 0.02 $\pm$ 0.01	\\
J1631.7$-$4756		& 31 		& 16 		& 9 		& 247.925 $\pm$ 0.023		& $-$47.944 $\pm$ 0.022		& 0.26 $\pm$ 0.02  $\pm$ 0.08	\\
J1633.0$-$4746		& 181 	& 146 	& 17 		& 248.259 $\pm$ 0.018		& $-$47.771 $\pm$ 0.025		& 0.61 $\pm$ 0.02 $\pm$ 0.12		\\
J1636.3$-$4731		& 71 		& 17 		& 8 		& 249.080 $\pm$ 0.020		& $-$47.522 $\pm$ 0.022		& 0.14 $\pm$ 0.02 $\pm$ 0.02	\\
J1652.2$-$4633$\dagger$		& 255 	& 212 	& 68 		& 253.055 $\pm$ 0.025		& $-$46.556 $\pm$ 0.022		& 0.72 $\pm$ 0.02 $\pm$ 0.04	\\
J1655.6$-$4738$\dagger$		& 46 		& 27 		& 2 		& 253.886 $\pm$ 0.030		& $-$47.638 $\pm$ 0.031		& 0.33 $\pm$ 0.03 $\pm$ 0.13	\\
J1713.7$-$3945		& 321 	& 255 	& 48 		& 258.433 $\pm$ 0.018		& $-$39.760 $\pm$ 0.019		& 0.55 $\pm$ 0.02 $\pm$ 0.01	\\
J1714.3$-$3823		& 139 	& 46 		& 44 		& 258.569 $\pm$ 0.021		& $-$38.391 $\pm$ 0.017		& 0.26 $\pm$ 0.02 $\pm$ 0.01	\\
J1745.8$-$3028$\dagger$ & 96 	         & 78         & 26 	& 266.453 $\pm$ 0.031		& $-$30.475 $\pm$ 0.028		& 0.53 $\pm$ 0.02 $\pm$ 0.26	(4/8) \\
J1800.6$-$2343		& 723	& 588	& 140	& 270.144 $\pm$ 0.022		& $-$23.716 $\pm$ 0.018		& 0.64 $\pm$ 0.01 $\pm$ 0.03	\\
J1804.8$-$2144		& 463	& 351	& 96 		& 271.197 $\pm$ 0.017		& $-$21.732 $\pm$ 0.017		& 0.38 $\pm$ 0.02 $\pm$ 0.01	\\
J1825.2$-$1359		& 240	& 235	& 30 		& 276.296 $\pm$ 0.035		& $-$13.992 $\pm$ 0.033		& 1.05 $\pm$ 0.02 $\pm$ 0.25	\\
J1834.8$-$0848		& 133	& 76		& 24		&278.694 $\pm$ 0.020		& $-$8.798 $\pm$ 0.022		& 0.29 $\pm$ 0.02 $\pm$ 0.01	\\
J1834.1$-$0706		& 110 	& 59 		& 29 		& 278.529 $\pm$ 0.018		& $-$7.109 $\pm$ 0.018		& 0.21 $\pm$ 0.02 $\pm$ 0.01		\\
J1836.5$-$0652		& 251 	& 207 	& 50 		& 279.143 $\pm$ 0.032		& $-$6.866 $\pm$ 0.034		& 0.54 $\pm$ 0.05 $\pm$ 0.06		\\
J1839.0$-$0704		& 117 	& 99 		& 45 		& 279.745 $\pm$ 0.027		& $-$7.067 $\pm$ 0.032		& 0.52 $\pm$ 0.02 $\pm$ 0.02	\\
J1839.4$-$0554		& 115 	& 104 	& 20 		& 279.856 $\pm$ 0.024		& $-$5.908 $\pm$ 0.025		& 0.41 $\pm$ 0.02 $\pm$ 0.05	\\
J1841.4$-$0514		& 157 	& 126 	& 15  	& 280.347	 $\pm$ 0.027		& $-$5.235 $\pm$ 0.025		& 0.47 $\pm$ 0.02 $\pm$ 0.01	\\
J1856.3+0122		& 232 	& 127 	& 68 		& 284.066 $\pm$ 0.023		& 1.369 $\pm$ 0.021			& 0.38 $\pm$ 0.02 $\pm$ 0.03		\\
J1857.8+0246		& 86 		& 65 		& 12 		& 284.449 $\pm$ 0.027		& 2.774 $\pm$ 0.042			& 0.61 $\pm$ 0.03 $\pm$ 0.06		\\
J1923.3+1408		& 349 	& 222 	& 67 		& 290.825	 $\pm$ 0.012		& 14.139 $\pm$ 0.014		& 0.29 $\pm$ 0.01 $\pm$ 0.01		\\
J2020.8+4026		& 338 	& 263 	& 51 		& 305.204 $\pm$ 0.020		& 40.443 $\pm$ 0.018		& 0.58 $\pm$ 0.01 $\pm$ 0.02		\\
J2026.1+4111		& 134 	& 125 	& 36 		& 306.534 $\pm$ 0.041		& 41.190 $\pm$ 0.036		& 1.37 $\pm$ 0.02 $\pm$ 0.26 (6/8)		\\
J2302.0+5855		& 54 		& 26 		& 16 		& 345.494 $\pm$ 0.026		& 58.920 $\pm$ 0.023		& 0.25 $\pm$ 0.02 $\pm$ 0.01		\\
\enddata
\tablecomments{Results of the maximum likelihood spatial fits for LAT-detected extended sources. Column 2 lists the TS of the source assuming it is spatially-extended with a disk spatial model whose position and extension are provided in columns 5, 6 (in Equatorial coordinates in J2000 epoch) and 7. Column 3 provides the TS$_{\rm ext}$ value which is twice the logarithm of the likelihood ratio of an extended to a point source, as defined in Section~\ref{subsection:SourceDetection}. The first error on the disk extension $\sigma$ is statistical and the second is systematic. The systematic error of three sources were computed using only a fraction of the 8 alternate IEMs since the likelihood maximization had convergence problems for the other IEMs. The number of alternate diffuse used is written into parentheses in column 7. Sources flagged with $\dagger$ are confused or contaminated by the diffuse background in complex regions.}
\end{deluxetable*}

\newpage
\begin{deluxetable*}{ccccc}
\tablecaption{Best-fit spectral parameters for the extended sources detected above 10 GeV\label{tab:spectra}}
\tablenum{2}
\tablehead{
\colhead{FGES Name} 	&
\colhead{TS$_{\rm curve}$} 	&
\colhead{Spectral} 	&
\colhead{Flux} 	&
\colhead{Spectral}\\
\colhead{} 	& \colhead{} 	& \colhead{Form} 	& \colhead{[$\times$~10$^{-10}$ cm$^{-2}$~s$^{-1}$]} 	& \colhead{index}
}
\startdata
J0427.2+5533			& 1 	& PL 	& $6.67 \pm 0.64 \pm 0.20$ 	& $1.91 \pm 0.09 \pm 0.02$	\\
J0537.6+2751			& 2 	& PL 	& $3.15 \pm 0.62 \pm 0.60$ 	& $2.10 \pm 0.17 \pm 0.05$ 	\\
J0617.2+2235			& 1 	& PL 	& $23.43 \pm 1.18 \pm 0.20$ 	& $2.55 \pm 0.08\pm 0.03$ 		\\
J0619.6+2229			& 3 	& PL 	& $4.30 \pm 0.69 \pm 0.09$ 	& $2.08 \pm 0.15 \pm 0.08$ \\
J0822.1$-$4253			& 1 	& PL 	& $6.59 \pm 0.57 \pm 0.09$ 	& $2.40 \pm 0.12 \pm 0.02$ \\
J0830.3$-$4453			& 1 	& PL 	& $1.89 \pm 0.34 \pm 0.09$ 	& $2.80 \pm 0.30 \pm 0.08$ \\
J0832.0$-$4549			& 1 	& PL 	& $2.72 \pm 0.48 \pm 0.12$ 	& $1.80 \pm 0.14 \pm 0.13$ \\
J0851.9$-$4620			& 8 	& PL 	& $17.23 \pm 1.14 \pm 1.10$ 	& $1.83 \pm 0.03 \pm 0.07$ \\
J1023.3$-$5747			& 1 	& PL 	& $4.23 \pm 0.50 \pm 0.32$ 	& $2.26 \pm 0.15 \pm 0.02$	\\
J1036.3$-$5834$\dagger$	& 11 	& PL 	& $29.11 \pm 1.88 \pm 1.77$ 	& $2.08 \pm 0.06 \pm 0.07$	\\
J1109.4$-$6115$\dagger$	& 2 	& PL 	& $10.58 \pm 1.02 \pm 0.63 $ 	& $2.14\pm 0.09 \pm 0.04$	\\
J1213.3$-$6240			& 1 	& PL 	& $3.37 \pm 0.45 \pm 0.07$ 	& $2.53 \pm 0.20 \pm 0.02$ \\
J1303.5$-$6313			& 5 	& PL 	& $3.52 \pm 0.50 \pm 0.51$ 	& $1.81 \pm 0.11 \pm 0.04$ \\
J1355.1$-$6420			& 4 	& PL 	& $1.78 \pm 0.32 \pm 0.20$ 	& $1.44 \pm 0.12 \pm 0.29$	\\
J1409.1$-$6121$\dagger$	& 2 	& PL 	& $16.45 \pm 1.24 \pm 1.02$		& $2.38 \pm 0.10 \pm 0.02$ \\
J1420.3$-$6047			& 3 	& PL 	& $3.19 \pm 0.43 \pm 0.68$ 	& $1.92 \pm 0.12 \pm 0.03$	\\
J1443.2$-$6227			& 2 	& PL 	& $2.46 \pm 0.30 \pm 0.09$ 	& $1.75 \pm 0.08 \pm 0.10$ \\
J1507.6$-$6228			& 1 	& PL 	& $2.28 \pm 0.28 \pm 0.36$ 	& $1.98 \pm 0.09 \pm 0.17$ \\
J1514.3$-$5910			& 3 	& PL 	& $7.69 \pm 0.50 \pm 0.17$ 	& $1.97 \pm 0.05 \pm 0.06$	\\
J1552.9$-$5610			& 2 	& PL 	& $5.95 \pm 0.51 \pm 0.03$ 	& $2.44 \pm 0.12 \pm 0.04$ \\
J1553.8$-$5325$\dagger$	& 4 	& PL 	& $11.75 \pm 1.01 \pm 0.24$ 	& $2.45 \pm 0.12 \pm 0.01$	\\
J1615.4$-$5153			& 5 	& PL 	& $9.88 \pm 0.79 \pm 1.26$ 	& $2.02 \pm 0.08 \pm 0.03$	\\
J1617.3$-$5054			& 2 	& PL 	& $14.70 \pm 1.06 \pm 0.32$ 	& $2.08 \pm 0.07 \pm 0.01$	\\
J1631.7$-$4756			& 2 	& PL 	& $4.19 \pm 0.84 \pm 0.37$ 	& $1.74 \pm 0.12 \pm 0.12$	\\
J1633.0$-$4746			& 2 	& PL 	& $18.51 \pm 0.14 \pm 0.37$ 	& $2.25 \pm 0.01 \pm 0.10$	\\
J1636.3$-$4731			& 1 	& PL 	& $4.30 \pm 0.17 \pm 1.44$ 	& $2.73 \pm 0.04 \pm 0.05$	\\
J1652.2$-$4633$\dagger$	& 1 	& PL  	& $11.95 \pm 0.97 \pm 1.74$ 	& $2.07 \pm 0.08 \pm 0.03$	\\
J1655.6$-$4738$\dagger$	& 6 	& PL 	& $2.11 \pm 0.41 \pm 0.11$ 	& $2.31 \pm 0.22 \pm 0.16$\\
J1713.7$-$3945			& 10 	& PL 	& $11.69 \pm 0.91 \pm 0.86$ 	& $1.85 \pm 0.07 \pm 0.02$	\\
J1714.3$-$3823			& 2	& PL 	& $6.08 \pm 0.68 \pm 0.70$ 	& $2.41 \pm 0.15 \pm 0.08$\\
J1745.8$-$3028$\dagger$ 	& 2 	& PL 	& $7.53 \pm 0.92 \pm 0.71$ 	& $2.09 \pm 0.13 \pm 0.03$ 	\\
J1800.6$-$2343			& 2	& PL 	& $27.47 \pm 1.08 \pm 0.71$ 	& $2.42 \pm 0.04 \pm 0.03$	\\
J1804.8$-$2144			& 7 	& PL 	& $15.55 \pm 0.62 \pm 0.60 $ 	& $1.99 \pm 0.04 \pm 0.05$	\\
J1825.2$-$1359			& 21	& LogP 	& $19.59 \pm 0.14 \pm 0.22 $ 	& $1.30 \pm 0.10 \pm 0.40$ \\
J1834.8$-$0848			& 4 	& PL 	& $7.43 \pm 0.79 \pm 0.12$ 	& $2.04 \pm 0.09 \pm 0.03$	\\
J1834.1$-$0706			& 1 	& PL 	& $5.37 \pm 0.66 \pm 0.78$ 	& $2.28 \pm 0.14 \pm 0.04$	\\
J1836.5$-$0652			& 9 	& PL 	& $17.98 \pm 1.31 \pm 1.72$ 	&	$2.10 \pm 0.07 \pm 0.03$ \\
J1839.0$-$0704			& 11 	& PL 	& $9.02 \pm 0.99 \pm 0.39 $ 	& $1.90 \pm 0.08 \pm 0.07$	\\
J1839.4$-$0554			& 1 	& PL 	& $8.39 \pm 0.94 \pm 0.81$ 	& $1.98 \pm 0.09 \pm 0.04$		\\
J1841.4$-$0514			& 4 	& PL 	& $9.48 \pm 0.91 \pm 0.92$ 	& $2.03 \pm 0.09 \pm 0.06$ \\
J1856.3+0122			& 1 	& PL 	& $10.44 \pm 0.92 \pm 1.33$ 	& $3.17 \pm 0.20 \pm 0.14$	\\
J1857.8+0246			& 2 	& PL 	& $7.83 \pm 1.01 \pm 1.39$ 	& $2.02 \pm 0.11 \pm 0.26$	\\
J1923.3+1408			& 1 	& PL 	& $12.52 \pm 0.97 \pm 0.97$ 	& $2.46 \pm 0.11 \pm 0.17$	\\
J2020.8+4026			& 1 	& PL 	& $13.22 \pm 0.81 \pm 0.29$ 	& $2.15 \pm 0.07 \pm 0.02$	\\
J2026.1+4111			& 7 	& PL 	& $15.80 \pm 1.32 \pm 5.73$ 	& $2.38 \pm 0.10 \pm 0.10$ 	\\
J2302.0+5855			& 1 	& PL 	& $1.26 \pm 0.25 \pm 0.04$ 	& $2.03 \pm 0.20 \pm 0.04$	\\
\enddata
\tablecomments{Results of the maximum likelihood spectral fits for
  LAT-detected extended sources. These results are obtained assuming
  the best Disk parameters reported in Table~\ref{tab:morpho}. Columns
  2, 4 and 5 report TS$_{\rm curve}$, the integrated flux and the photon index of the source fit in the energy range from 10 GeV to
  2 TeV. The first error on the integrated flux and photon index is
  statistical and the second is systematic. Column 3 lists the
  spectral form used (PL = Power-Law, LogP = LogParabola). J1825.2$-$1359 is the only
  source modeled with a LogP and its associated beta value is $0.27
  \pm 0.05 \pm 0.07$. Sources flagged with $\dagger$ are confused or contaminated by the diffuse background in complex regions.}
\end{deluxetable*}

\begin{deluxetable*}{cccccc}
\tablecaption{Best-fitting morphological and spectral parameters for the systematic study using a Gaussian fit\label{tab:gauss}}
\tablenum{3}
\tablehead{
\colhead{FGES Name} &
\colhead{Right Ascension} &
\colhead{Declination} &
\colhead{$\sigma$} &
\colhead{Flux} &
\colhead{Spectral Index}  \\
\colhead{} & \colhead{{\bf (\degrees)}} & \colhead{{\bf (\degrees)}} & \colhead{(\degrees)} & \colhead{[$\times$~10$^{-10}$ cm$^{-2}$~s$^{-1}$]} & \colhead{}
}
\startdata
J0427.2+5533		& 66.95 $\pm$ 0.11		& 55.35 $\pm$ 0.10		& 0.92 $\pm$ 0.03  & 8.16 $\pm$ 0.76 & 1.93 $\pm$ 0.08 		\\
J0537.6+2751		& 84.41 $\pm$ 0.18		& 27.76 $\pm$ 0.15		& 0.71 $\pm$ 0.09 & 2.74 $\pm$ 0.53 & 2.03 $\pm$ 0.09		\\
J0617.2+2235		& 94.31 $\pm$ 0.01		& 22.57 $\pm$ 0.01		& 0.18$\pm$ 0.01 & 24.05 $\pm$ 1.26 & 2.56 $\pm$ 0.08		\\
J0619.6+2229		& 94.56 $\pm$ 0.07		& 22.53 $\pm$ 0.07		& 0.54 $\pm$ 0.05 & 4.45 $\pm$ 0.85 & 2.01 $\pm$ 0.15	\\
J0822.1$-$4253		& 125.65 $\pm$ 0.03		& $-$42.88 $\pm$ 0.02		& 0.24 $\pm$ 0.02 & 6.97 $\pm$ 0.58 & 2.38 $\pm$ 0.06		\\
J0830.3$-$4453		& 127.65 $\pm$ 0.04		& $-$44.88 $\pm$ 0.04		& 0.17 $\pm$ 0.03 & 2.14 $\pm$ 0.54 & 2.85 $\pm$ 0.06	\\
J0832.0$-$4549		& 127.97 $\pm$ 0.07		& $-$45.81 $\pm$ 0.07		&  0.48 $\pm$ 0.04 & 4.12 $\pm$ 0.57 & 1.90 $\pm$ 0.05	\\
J0851.9$-$4620		& 132.86 $\pm$ 0.05		& $-$46.34 $\pm$ 0.05		& 0.72 $\pm$ 0.03 & 20.29 $\pm$ 1.51 & 1.85 $\pm$ 0.02	\\
J1023.3$-$5747		& 155.84 $\pm$ 0.02		& $-$57.75 $\pm$ 0.02		& 0.16 $\pm$ 0.02 & 4.86 $\pm$ 0.53 & 2.23 $\pm$ 0.10	\\
J1036.3$-$5834$\dagger$ & 158.94 $\pm$ 0.06		& $-$58.77 $\pm$ 0.06		& 1.57 $\pm$ 0.06 & 36.27 $\pm$ 2.57 & 2.10 $\pm$ 0.04	\\
J1109.4$-$6115$\dagger$ & 166.90 $\pm$ 0.08		& $-$61.20 $\pm$ 0.07		& 0.88 $\pm$ 0.05 & 14.46 $\pm$ 1.17 & 2.17 $\pm$ 0.04	\\
J1213.3$-$6240		& 183.28 $\pm$ 0.03		& $-$62.69 $\pm$ 0.03		& 0.18 $\pm$ 0.02 & 3.83 $\pm$ 0.50 & 2.60 $\pm$ 0.16 		\\
J1303.5$-$6313		& 195.84 $\pm$ 0.03		& $-$63.20 $\pm$ 0.03		& 0.19 $\pm$ 0.02 & 4.02 $\pm$ 0.55 & 1.81 $\pm$ 0.10		\\
J1355.1$-$6420		& 208.75 $\pm$ 0.03		& $-$64.44 $\pm$ 0.03		& 0.22 $\pm$ 0.03 & 1.80 $\pm$ 0.24 & 1.44 $\pm$ 0.03 		\\
J1409.1$-$6121$\dagger$ & 212.37 $\pm$ 0.030		& $-$61.31 $\pm$ 0.03		& 0.51 $\pm$ 0.02 & 20.63 $\pm$ 1.52 & 2.36 $\pm$ 0.09		\\
J1420.3$-$6047		& 215.07 $\pm$ 0.02		& $-$60.77 $\pm$ 0.02		& 0.11 $\pm$ 0.02 & 4.54 $\pm$ 0.26 & 1.99 $\pm$ 0.08		\\
J1443.2$-$6227		& 220.80 $\pm$ 0.03		& $-$62.41 $\pm$ 0.03		& 0.19 $\pm$ 0.02 & 2.46 $\pm$ 0.21 & 1.72 $\pm$ 0.04	\\
J1507.6$-$6228		& 226.92 $\pm$ 0.04		& $-$62.44 $\pm$ 0.04		& 0.25 $\pm$ 0.04 & 2.41 $\pm$ 0.27 & 2.03 $\pm$ 0.03	\\
J1514.3$-$5910		& 228.55 $\pm$ 0.01		& $-$59.17 $\pm$ 0.01		& 0.13 $\pm$ 0.01 & 7.68 $\pm$ 0.51 & 1.97 $\pm$ 0.02	\\
J1552.9$-$5610		& 238.18 $\pm$ 0.02		& $-$56.18 $\pm$ 0.02		& 0.14 $\pm$ 0.01 & 6.01 $\pm$ 0.51 & 2.43 $\pm$ 0.11	\\
J1553.8$-$5325$\dagger$ & 238.50 $\pm$ 0.03		& $-$53.44 $\pm$ 0.03		& 0.35 $\pm$ 0.02 & 15.43 $\pm$ 1.22 & 2.41 $\pm$ 0.10	\\
J1615.4$-$5153		& 243.77 $\pm$ 0.03		& $-$51.86 $\pm$ 0.03		& 0.34 $\pm$ 0.03 & 12.72 $\pm$ 0.98 & 2.00 $\pm$ 0.07 	\\
J1617.3$-$5054		& 244.27 $\pm$ 0.03		& $-$50.93 $\pm$ 0.02		& 0.30 $\pm$ 0.01 & 17.22 $\pm$ 1.22 & 2.12 $\pm$ 0.07	\\
J1631.7$-$4756		& 247.96 $\pm$ 0.03		& $-$47.98 $\pm$ 0.03		& 0.14 $\pm$ 0.03 & 5.61 $\pm$ 0.34 & 1.82 $\pm$ 0.12	\\
J1633.0$-$4746		& 248.40 $\pm$ 0.03		& $-$47.71 $\pm$ 0.03		& 0.44 $\pm$ 0.03 & 22.13 $\pm$ 1.77 & 2.28 $\pm$ 0.03	\\
J1636.3$-$4731		& 250.16 $\pm$ 0.04		& $-$46.57 $\pm$ 0.04		& 0.05 $\pm$ 0.01 & 5.22 $\pm$ 0.45 & 2.02 $\pm$ 0.02	\\
J1652.2$-$4633$\dagger$ & 253.09 $\pm$ 0.01		& $-$46.50 $\pm$ 0.01		& 0.48 $\pm$ 0.03 & 15.32 $\pm$ 1.02 & 2.08 $\pm$ 0.02 	\\
J1655.6$-$4738$\dagger$ & 253.93 $\pm$ 0.06		& $-$47.65 $\pm$ 0.06		& 0.29 $\pm$ 0.06 & 2.48 $\pm$ 0.45 & 2.31 $\pm$ 0.07 	\\
J1713.7$-$3945		& 258.39 $\pm$ 0.03		& $-$39.82 $\pm$ 0.03		& 0.41 $\pm$ 0.01 & 14.11 $\pm$ 1.14 & 1.91 $\pm$ 0.06	\\
J1714.3$-$3823		& 258.57 $\pm$ 0.02		& $-$38.42 $\pm$ 0.02		& 0.14 $\pm$ 0.02 & 6.46 $\pm$ 0.71 & 2.42 $\pm$ 0.11 	\\
J1745.8$-$3028$\dagger$ & 266.52 $\pm$ 0.04		& $-$30.43 $\pm$ 0.04		& 0.26 $\pm$ 0.02 & 7.29 $\pm$ 0.87 & 2.09 $\pm$ 0.06 \\
J1800.6$-$2343		& 270.17 $\pm$ 0.02		& $-$23.73 $\pm$ 0.02		& 0.37 $\pm$ 0.02 & 30.55 $\pm$ 1.17 & 2.41 $\pm$ 0.04 	\\
J1804.8$-$2144		& 271.20 $\pm$ 0.02		& $-$21.74 $\pm$ 0.02		& 0.24 $\pm$ 0.02 & 17.76 $\pm$ 0.96 & 2.01 $\pm$ 0.02	\\
J1825.2$-$1359		& 276.33 $\pm$ 0.05		& $-$13.97 $\pm$ 0.05		& 0.79 $\pm$ 0.04 &  29.45 $\pm$ 1.99 & 1.54 $\pm$ 0.08	\\
J1834.8$-$0848		& 278.67 $\pm$ 0.02		& $-$8.78 $\pm$ 0.03		& 0.15 $\pm$ 0.02 & 8.15 $\pm$ 0.84 & 2.04 $\pm$ 0.06 \\
J1834.1$-$0706		& 278.53 $\pm$ 0.02		& $-$7.11 $\pm$ 0.02		& 0.15 $\pm$ 0.02 & 5.10 $\pm$ 0.77 & 2.39 $\pm$ 0.11		\\
J1836.5$-$0652		& 279.10 $\pm$ 0.03		& $-$6.87 $\pm$ 0.03		& 0.38 $\pm$ 0.02 & 23.30 $\pm$ 1.73 & 2.12 $\pm$ 0.05 		\\
J1839.0$-$0704		& 279.75 $\pm$ 0.04		& $-$7.04 $\pm$ 0.04		& 0.37 $\pm$ 0.03 & 9.13 $\pm$ 1.18  & 1.94 $\pm$ 0.06 	\\
J1839.4$-$0554		& 279.90 $\pm$ 0.03		& $-$5.90 $\pm$ 0.03		& 0.25 $\pm$ 0.02 & 9.02 $\pm$ 1.05 & 2.03 $\pm$ 0.06	\\
J1841.4$-$0514		& 280.31	 $\pm$ 0.04		& $-$5.22 $\pm$ 0.03		& 0.31 $\pm$ 0.03 & 10.90 $\pm$ 1.09 & 2.04 $\pm$ 0.07	\\
J1856.3+0122		        & 283.99 $\pm$ 0.02		& 1.42 $\pm$ 0.02			& 0.21 $\pm$ 0.02 & 11.17 $\pm$ 0.98 & 3.17 $\pm$ 0.20		\\
J1857.8+0246		        & 284.40 $\pm$ 0.04		& 2.80 $\pm$ 0.04			& 0.32 $\pm$ 0.03 & 8.25 $\pm$ 0.98 & 2.02 $\pm$ 0.07 		\\
J1923.3+1408		        & 290.81	 $\pm$ 0.01		& 14.14 $\pm$ 0.01		& 0.17 $\pm$ 0.01 & 13.17 $\pm$ 1.05 & 2.54 $\pm$ 0.12		\\
J2020.8+4026		        & 305.21 $\pm$ 0.02		& 40.46 $\pm$ 0.02		& 0.35 $\pm$ 0.01 & 16.34 $\pm$ 1.03 & 2.21 $\pm$ 0.03 		\\
J2026.1+4111		        & 307.16 $\pm$ 0.07		& 41.45 $\pm$ 0.07		& 1.29 $\pm$ 0.06 & 35.95 $\pm$ 2.59  & 2.40 $\pm$ 0.03	\\
J2302.0+5855		        & 345.53 $\pm$ 0.03		& 58.89 $\pm$ 0.03		& 0.14 $\pm$ 0.02 & 1.32 $\pm$ 0.27 & 2.05 $\pm$ 0.17	\\
\enddata
\tablecomments{Results of the maximum likelihood spatial and spectral fits for LAT-detected extended sources using a Gaussian spatial model. The position and extension of the Gaussian are provided in columns 2, 3 (in Equatorial coordinates in J2000 epoch) and 4. The error quoted is only statistical. The sigma value for a disk is expected to be a factor of 1.85 larger than the sigma for a 2D Gaussian fit to the same source \citep{2012ApJ...756....5L}. J1825.2$-$1359 is the only source modeled with a LogP and its associated beta value is $0.23 \pm 0.05$. Sources flagged with $\dagger$ are confused or contaminated by the diffuse background in complex regions.}
\end{deluxetable*}

\acknowledgments
The \textit{Fermi} LAT Collaboration acknowledges generous ongoing support
from a number of agencies and institutes that have supported both the
development and the operation of the LAT as well as scientific data analysis.
These include the National Aeronautics and Space Administration and the
Department of Energy in the United States, the Commissariat \`a l'Energie Atomique
and the Centre National de la Recherche Scientifique / Institut National de Physique
Nucl\'eaire et de Physique des Particules in France, the Agenzia Spaziale Italiana
and the Istituto Nazionale di Fisica Nucleare in Italy, the Ministry of Education,
Culture, Sports, Science and Technology (MEXT), High Energy Accelerator Research
Organization (KEK) and Japan Aerospace Exploration Agency (JAXA) in Japan, and
the K.~A.~Wallenberg Foundation, the Swedish Research Council and the
Swedish National Space Board in Sweden.
 
Additional support for science analysis during the operations phase is gratefully
acknowledged from the Istituto Nazionale di Astrofisica in Italy and the Centre
National d'\'Etudes Spatiales in France. This work performed in part under DOE
Contract DE-AC02-76SF00515.

\software{Fermi Science Tools (v10r01p01), $\mathtt{pointlike}$ \citep{2010PhDT.......147K}}.

\appendix

\section{Systematic cross-check with a secondary pipeline}
\label{appen:xCheck}

This paper used two analysis pipelines similar to those employed in \cite{2016ApJS..222....5A}. The primary one is presented in $\S$\ref{subsection:SourceDetection}. Both methods implemented $\mathtt{pointlike}$, but each made slightly different choices about how to construct the region model and update the spectral and spatial parameters of surrounding sources as new sources were added or removed in each field. The pipelines reached a highly compatible representation of sources along the Galactic plane that accounted for the presence of extended sources. The only two sources with significant disagreement were rejected from the list presented and are discussed below. The use of two independent analysis pipelines provided detailed crosschecks of a large-scale, multi-step analysis and determined how algorithm choices impacted the final source model for an ROI. The two pipelines followed a similar procedure with the following exceptions:

\begin{itemize}

\item{The secondary pipeline considered a single row of 72 partially overlapping ROIs of radius 10$^\circ$, centered on b = 0$^\circ$, whereas the primary pipeline included two additional overlapping rows centered on b = $\pm$5$^\circ$ (see right diagram of Figure \ref{fig:pipeline}) creating 216 total ROIs.}

\item{Only 3FGL sources that were identified \citep[as defined in][]{2015ApJS..218...23A} with a multi-wavelength counterpart were retained in the initial region models used in the secondary pipeline, i.e. unassociated sources were not included. Extended sources identified as an SNR or PWN, as well as the Cygnus cocoon, were also excluded from the initial region models. The primary pipeline instead included all sources listed in the 3FGL catalog in the initial models and modified the spatial templates for extended sources to be a compatibly-sized uniform disk if the 3FGL template was not a uniform disk.}

\item{The secondary pipeline included an initial fit of all spectral parameters within the full 10$^\circ$ radius region. All following iterations left spectral parameters for sources within 5$^\circ$ of the newly added source free with all others fixed, the same as the primary pipeline. The primary pipeline could forego that initial step because it began with a more complete model, as described above, and as the final step in constructing the source model adjusted the parameters for sources appearing in multiple ROIs by using the fit from the one with the closest center. Many sources within the 10$^\circ$ radius b = 0$^\circ$ ROIs but lying beyond 5$^\circ$ of the center,  lie within 5$^\circ$ of an ROI centered on b = $\pm$5$^\circ$ (see Figure~\ref{fig:pipeline}).
}

\item{After a new point or extended source was added to the model, the spatial parameters of any previously added extended sources were refit iteratively, starting with the highest TS extended source, before creating a new TS map and continuing the iteration. The primary pipeline instead refit sources during the iteration only if a source TS fell below threshold and was removed from the model.}

\item{To finalize the source model, any sources with $\rm {TS <16}$ were removed from the ROI iteratively, starting with the lowest TS source, and all sources within 5$^\circ$ of the removed source were refit on each iteration. The primary pipeline removed sources with $\rm {TS <16}$ at each step.}

\end{itemize}

The fact that the two pipelines agree well on all sources presented here is very reassuring. However, it is clear that the secondary pipeline uses many more iterations, and therefore more computing resources, for each region because of excluding a number of 3FGL sources from the initial model that in many cases return in later iterations. The extended sources are refit each time additional sources are added to the model, creating an additional computational burden that influenced the choice to select 72 regions instead of 216 as is done in the primary pipeline. Consequentially, the primary pipeline covers the Galactic plane a little more thoroughly (see Figure~\ref{fig:pipeline}), making it the preferred analysis for this work.

There were 2 sources rejected from the list presented here due to the disagreement between the 2 pipelines. Both were located at the edge of search regions along the Galactic plane. One, coincident with the Cygnus loop (l = 73.98, b = -8.56) with a disk radius of $1.6^{\circ}$ and a $TS_{\rm ext}$ value of 21, was detected by the primary pipeline but was not detected by the secondary pipeline. This can be explained by the large offset of this source with respect to the Galactic plane causing it to not be included in the search performed by the secondary pipeline. The second rejected source was only detected by the secondary pipeline with a disk radius of $0.06^{\circ}$ whereas the primary pipeline found a point source at the same position (l = 276.12, b = -7.04), coincident with the \emph{Fermi}-LAT source 3FGL J0904.8$-$5734 (associated with PKS 0903-57). It seems very likely that the extension estimates for these two sources are incorrect and affected by their location at the edge of the ROI for each pipeline. This explains why they were rejected from the final list.

\bibliographystyle{hapj}
\bibliography{bibliography.bib}

\end{document}